\documentclass[12pt,preprint]{aastex}

\begin{document}
\title{A $\sim$7.5 Earth-Mass Planet Orbiting the Nearby Star, GJ~876}

\bigskip
\author{
Eugenio J. Rivera\altaffilmark{2,3,4},
Jack J. Lissauer\altaffilmark{3},
R. Paul Butler\altaffilmark{4},
Geoffrey W. Marcy\altaffilmark{5},
Steven S. Vogt\altaffilmark{2},
Debra A. Fischer\altaffilmark{6},
Timothy M. Brown\altaffilmark{7},
Gregory Laughlin\altaffilmark{2},
Gregory W. Henry\altaffilmark{8,9}
}

\email{rivera@ucolick.org}

\altaffiltext{1}{Based on observations obtained at the W.M. Keck Observatory,
which is operated jointly by the University of California and the California
Institute of Technology.}

\altaffiltext{2}{UCO/Lick Observatory, University of California at Santa Cruz,
Santa Cruz, CA 95064}

\altaffiltext{3}{NASA/Ames Research Center, Space Science and Astrobiology
Division, MS 245-3, Moffett Field, CA 94035}

\altaffiltext{4}{Department of Terrestrial Magnetism, Carnegie Institution
of Washington, 5241 Broad Branch Road NW, Washington DC, 20015-1305}

\altaffiltext{5}{Department of Astronomy, University of California,
Berkeley, CA 94720}

\altaffiltext{6}{Department of Physics and Astronomy, San Francisco
State University, San Francisco, CA 94132}

\altaffiltext{7}{High Altitude Observatory, National Center for
Atmospheric Research, P.O. Box 3000, Boulder, CO 80307}

\altaffiltext{8}{Center of Excellence in Information Systems,
Tennessee State University, 3500 John A. Merritt Blvd., Box 9501, Nashville, TN 37209}

\altaffiltext{9}{Department of Physics and Astronomy,
Vanderbilt University, Nashville, TN  37235}

\begin{abstract}

High precision, high cadence radial velocity monitoring over the past 8 years
at the W. M. Keck Observatory reveals evidence for a third planet orbiting the
nearby (4.69 pc) dM4 star GJ~876.  The residuals of three-body Newtonian fits,
which include GJ~876 and Jupiter mass companions b and c, show significant
power at a periodicity of 1.9379 days.  Self-consistently fitting the radial
velocity data with a model that includes an additional body with this period
significantly improves the quality of the fit.  These four-body (three-planet)
Newtonian fits find that the minimum mass of companion ``d'' is
$m\sin{i}=5.89\,\pm\,0.54\,M_{\oplus}$ and that its orbital period is
$1.93776\,(\pm\,7\times10^{-5})$ days. Assuming coplanar orbits, an inclination
of the GJ~876 planetary system to the plane of the sky of $\sim50^{\circ}$
gives the best fit. This inclination yields a mass for companion d of
$m=7.53\,\pm\,0.70\,M_{\oplus}$, making it by far the lowest mass companion yet
found around a main sequence star other than our Sun.  Precise photometric
observations at Fairborn Observatory confirm low-level brightness variability
in GJ~876 and provide the first explicit  determination of the star's 96.7-day
rotation period.  Even higher precision short-term photometric measurements
obtained at Las Campanas imply that planet d does not transit GJ~876.

\end{abstract}


\section{Introduction}
\label{intro}

GJ~876 (HIP 113020) is the lowest mass star currently known to harbor planets.
The first companion discovered, ``b,'' was announced by Marcy et al. (1998) and
Delfosse et al. (1998).  They found that this companion had an orbital period,
$P_b$, of $\sim 61$ days and a minimum mass ($m_b\sin{i_b}$) of
$\sim 2.1\,M_{\rm Jup}$ and that it produced a reflex barycentric velocity
variation of its dM4 parent star of amplitude $K_b \sim 240$ m\,s$^{-1}$.
After 2.5 more years of continued Doppler monitoring, Marcy et al. (2001)
announced the discovery of a second companion, ``c.''  This second companion
has an orbital period, $P_c$, of $\sim$ 30 days,
$m_c\sin{i_c} \sim 0.56\,M_{\rm Jup}$, and $K_c \sim 81$ m\,s$^{-1}$.  As a
result of fitting the radial velocity data with a model with two
non-interacting companions, the fitted parameters for companion b were
different, with the most significant change in $K_b$ (and correspondingly
$m_b\sin{i_b}$), which dropped from 240 m\,s$^{-1}$ to 210 m\,s$^{-1}$.

\setcounter{footnote}{9}

Marcy et al. (2001) noted that although a model with two planets on unperturbed
Keplerian orbits produces a very significantly improved fit to the radial
velocity data by dramatically reducing both the $\sqrt{\chi_{\nu}^2}$ and the
RMS of the fit\footnote{The $\chi^2$ is the sum of the squares of the
differences between the data points and model points at each observed epoch
divided by the corresponding uncertainties of the measurements.  The reduced
chi-squared, $\chi_{\nu}^2$, is that quantity divided by the number of degrees
of freedom (the total number of observations minus the number of fitted
parameters).  The RMS is the root-mean-square of the velocity residuals after
the model fit has been subtracted from the data.  Note that $\chi^2$ has no
units, and  RMS has units of m\,s$^{-1}$ in this case.}, these two statistics
were still relatively large.  A $\sqrt{\chi_{\nu}^2}$ of $\sim$ 1.0 is expected
for a model that is a ``good" fit to the data assuming normally (Gaussian)
distributed errors.  Additionally, dynamical simulations based on this model
showed that the system's stability is strongly dependent on the starting epoch,
which is used to determine the initial positions of the planets for the
integrations.  This indicated that the mutual perturbations among the planets
are substantial on orbital timescales (Marcy et al. 2001).  Laughlin \&
Chambers (2001) and Rivera \& Lissauer (2001) independently developed
self-consistent ``Newtonian'' fitting schemes which incorporate the mutual
perturbations among the planets in fitting the radial velocity data.  Nauenberg
(2002) developed a similar method which additionally gives a lower limit on the
mass of the star;  using the radial velocity data from Marcy et al. (2001), he
found the mass of GJ~876 to be $M_{\star}>0.3\,M_{\odot}$.  This dynamical
modeling resulted in a substantially improved fit to the radial velocity data.

Laughlin et al. (2005) provide an updated analysis of the GJ~876 planetary
system in which they perform 3-body (two-planet) fits to a radial velocity data
set which includes 16 old observations taken at Lick observatory and
observations taken at the Keck observatory up to the end of the 2003 observing
season.  In their fits, they have assumed a stellar jitter of 6 m\,s$^{-1}$.
They found that the two jovian-mass planets are deeply in both the 2:1 mean
motion resonance and in an apsidal resonance in which the longitudes of
periastron remain nearly aligned.  All three resonant angles librate with small
amplitudes, which argues for a dissipative history of differential migration
for the giant planets.  Additionally, they were able to constrain the
inclination of the coplanar system to the plane of the sky to be
$i > 20^{\circ}$.  Finally, they examined the possibility that if companion c
is not transiting now, transits might occur in the future for non-coplanar
configurations with modest mutual inclinations.

In this paper, we describe the results of a more detailed analysis using a new
radial velocity data set.  Note that most of the fits presented in this work do
not take stellar jitter (which is found to be $\lesssim$ 1.5 m\,s$^{-1}$, see
Section \ref{3plcf}) into account.  In Section \ref{obs}, we present the new
velocities and describe the procedures which resulted in significant
improvements in the precision of these velocities.  In Section \ref{2plcf}, we
incorporate the techniques from Laughlin et al. (2005) to determine the
uncertainties in the parameters from two-planet fits.  In Section
\ref{Res-2pl}, we present a periodogram analysis of the residuals to the
two-planet fit, which suggests the presence of a third companion to GJ~876,
with a period of 1.94 days.  In Section \ref{3plcf}, we present the results
from three-planet fits, which provide estimates of the actual masses of
companions b and c as well as $m_d \sin{i_d}$ of the small inner planet.  The
residuals of the 2-planet fit also show significant power at 2.0548 days; as
discussed in Section \ref{Aliasing}, we have demonstrated that this second
period is an alias of the 1.9379-day period.  Section \ref{photometry} presents
the results of long-term photometric monitoring of GJ~876.  In Section
\ref{transits}, we show that the third companion was not transiting in 2003. We
discuss some interesting aspects of the third companion in Section \ref{dis}.
Finally, we end with a summary of our results and our conclusions.

\section{Radial Velocity Observations}
\label{obs}

The stellar characteristics of GJ~876 (M4 V) have been described previously in
Marcy et al. (1998) and Laughlin et al. (2005).  It has a Hipparcos distance of
4.69 pc (Perryman et al. 1997).  From its distance and the bolometric
correction of Delfosse et al. (1998), its luminosity is 0.0124 $L_{\odot}$.  As
in previous studies, we adopt a stellar mass of $0.32\,M_{\odot}$ and a radius
of $0.3\,R_{\odot}$ based on the mass-luminosity relationship of Henry \&
McCarthy (1993).  We do not incorporate uncertainties in the star's mass
($0.32\,\pm\,0.03\,M_{\odot}$) into the uncertainties in planetary masses and
semi-major axes quoted herein.  The age of the star is roughly in the range
1\,--\,10 Gyr (Marcy et al. 1998).

\begin{figure}
\includegraphics[scale=0.775]{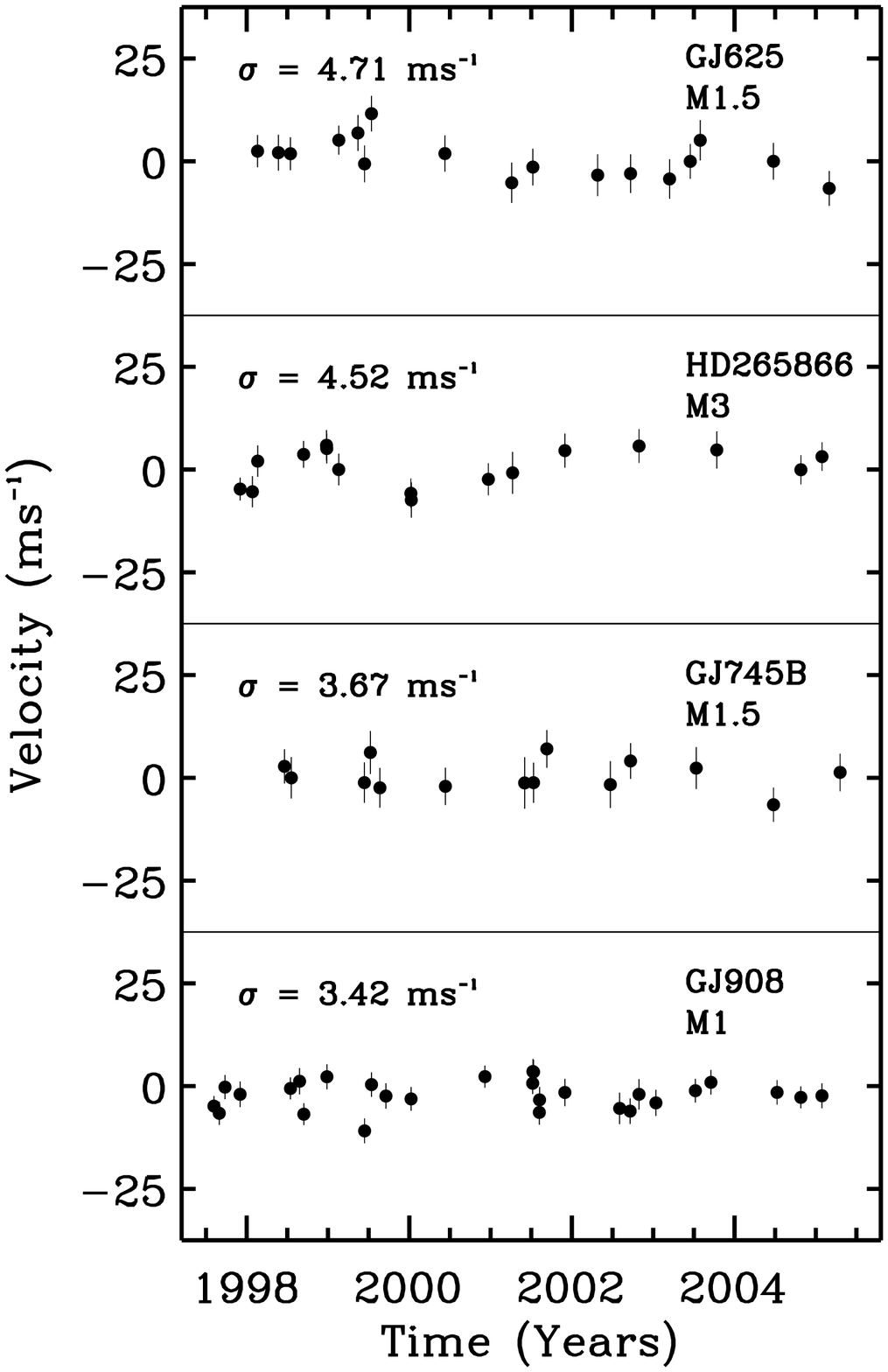}
\caption{
Radial velocity versus time for Keck M dwarfs spanning 7 years.  The Keck-HIRES
system achieves precision of 3\,--\,5 m\,s$^{-1}$ for M dwarfs brighter than V\,=\,11.
}
\label{keck_stable}
\end{figure}

We searched for Doppler variability using repeated, high resolution spectra
with resolving power R $\approx 70000$, obtained with the Keck/HIRES
spectrometer (Vogt et al. 1994).  The Keck spectra span the wavelength range
from 3900\,--\,6200 \AA.  An iodine absorption cell provides wavelength
calibration and the instrumental profile from 5000 to 6000 \AA \ (Marcy \&
Butler 1992, Butler et al. 1996).  Typical signal-to-noise ratios are 100 per
pixel for GJ~876.  At Keck we routinely obtain Doppler precision of
3\,--\,5 m\,s$^{-1}$ for V\,=\,10 M dwarfs, as shown in Figure
\ref{keck_stable}.  A different set of 4 stable Keck M dwarfs is shown in
Figure 2 of Butler et al. (2004).  The variations in the observed radial
velocities of these stars can be explained by the internal uncertainties in the
individual data points; thus, there is no evidence that any of these stars
possess planetary companions.  Exposure times for GJ~876 and other V\,=\,10
M dwarfs are typically 8 minutes.

The internal uncertainties in the velocities are judged from the velocity
agreement among the approximately 400 2-\AA\, chunks of the echelle spectrum,
each chunk yielding an independent Doppler shift.  The internal velocity
uncertainty of a given measurement is the uncertainty in the mean of the
$\sim 400$ velocities from one echelle spectrum.

We present results of N-body fits to the radial velocity data taken at the
W. M. Keck telescope from June 1997 to December 2004.  The 155 measured radial
velocities are listed in Table \ref{velocities}.  The median of the
uncertainties is 4.1 m\,s$^{-1}$.  Comparison of these velocities with those
presented in Laughlin et al. (2005) shows significant changes (typically
3\,--\,10 m\,s$^{-1}$) in the velocities at several observing epochs.

The changes in the measured velocities are a result of a more sophisticated
modeling of the spectrum at sub-pixel levels and of key improvements in various
instrumental idiosyncrasies. The previous HIRES CCD, installed at first-light
in 1993, had 1) relatively large (24 $\mu$m) pixels, 2) a convex surface, 3)
excessive charge diffusion in the CCD substrate, which broadened the detector's
point spread function (PSF), and 4) a subtle signal-dependent non-linearity in
charge transfer efficiency (CTE). These combined effects had been limiting our
radial velocity precision with HIRES to about 3 m\,s$^{-1}$ since 1996. In
August 2004, the old CCD was replaced by a new 3-chip mosaic of MIT-Lincoln
Labs CCDs. These devices provided a very flat focal plane (improving the
optical PSF), a finer pixel pitch (which improved our sub-pixel modeling), and
more spectral coverage per exposure. The MIT-LL devices also are free of
signal-dependent CTE non-linearities and have a much lower degree of charge
diffusion in the CCD substrate (which improved the detector PSF). We also
switched into higher cadence mode in October, 2004, observing 3 times per night
and for as many consecutive nights as telescope scheduling would allow.
Additionally, toward the end of 2004, a high signal-to-noise template of GJ~876
was obtained. All Keck data were then re-reduced using the improved Doppler
code together with the new high S/N template and the higher cadence 2004
observations.  As a result of the improvements, the two- (and three-) planet
fits presented here for this system are significantly improved over previous
N-body fits.

\section{Two-Planet Coplanar Fits}
\label{2plcf}

We first performed self-consistent 2-planet fits in which we assumed that the
orbits of both companions ``b'' and ``c'' are coplanar and that this plane
contains the line of sight ($i_b=i_c=90^{\circ}$).  These are fits to all the
155 Keck radial velocities listed in Table \ref{velocities}.  All fits were
obtained with a Levenberg-Marquardt minimization algorithm (Press et al. 1992)
driving an N-body integrator.  This algorithm is a more general form of the
algorithms used in Laughlin \& Chambers (2001) and Rivera \& Lissauer (2001).
All fits in this work are for epoch JD 2452490, a time near the center of the
155 radial velocity measurements.  We fitted for 10+1 parameters; 10 of these
parameters are related to the planetary masses and orbital elements: the
planetary masses ($m$), the semi-major axes ($a$), eccentricities ($e$),
arguments of periastron ($\omega$), and mean anomalies ($M$) of each body, and
1 parameter is for the radial velocity offset, representing the center-of-mass
motion of the GJ~876 system relative to the barycenter of our Solar System and
arbitrary zero-point of the velocities.

\begin{figure}
\includegraphics[angle=-90,scale=0.61]{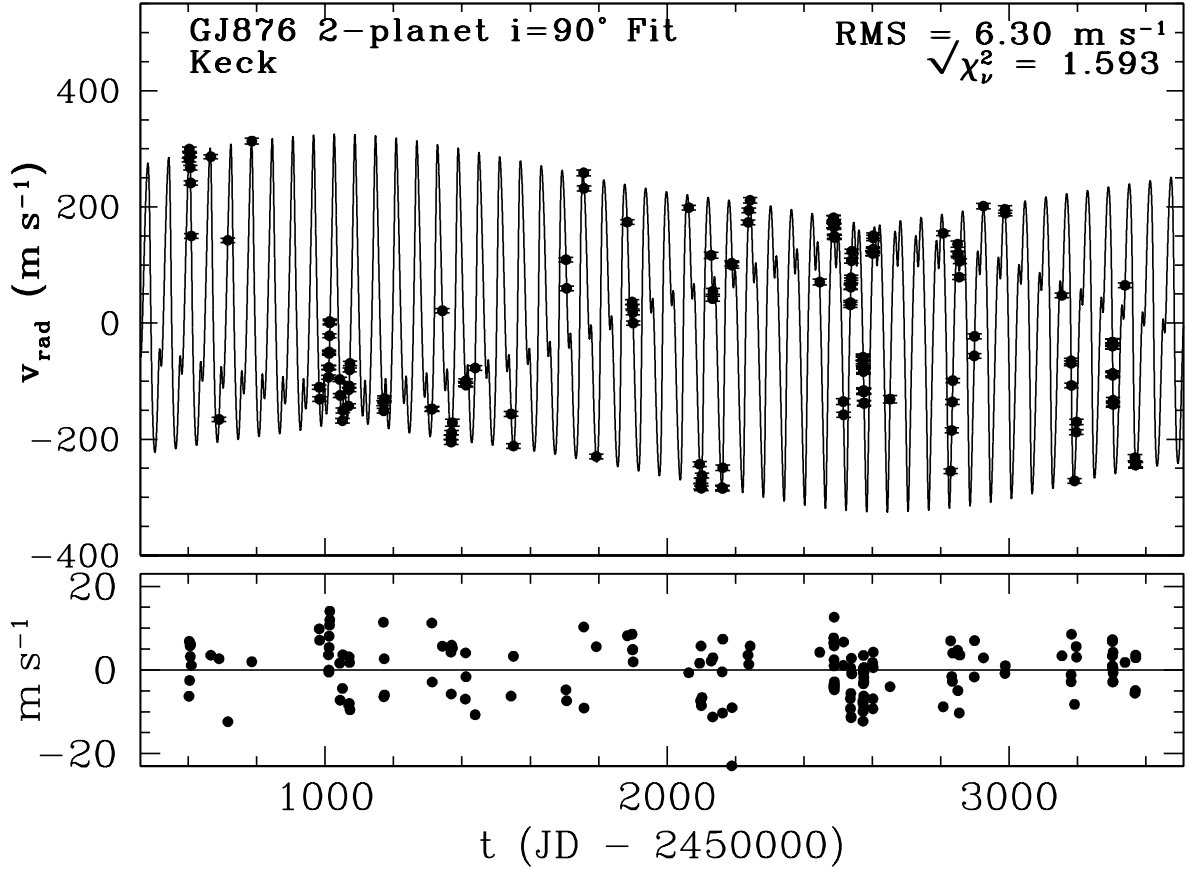}
\caption{
Top: Velocity versus time for GJ~876 with model radial velocity (solid line)
generated from the nominal self-consistent, coplanar, $i=90^{\circ}$,
three-body (two-planet) fit to the observed radial velocities.  The observed
velocities (from Table \ref{velocities}) are shown as small solid points with
vertical error bars corresponding to the instrumental uncertainties.
Bottom: Residuals to the orbital fit.
}
\label{2mf=1.00}
\end{figure}

We first obtained a nominal 2-planet fit to the actual 155 observed velocities.
Figure \ref{2mf=1.00} shows the model radial velocity (solid line) generated
from this nominal 2-planet fit, along with the actual observed velocities
(solid points with vertical error bars); the residuals are shown in the lower
portion.  Table \ref{2plparam} shows the best fitted parameters, which are
similar to those obtained by Laughlin et al. (2005).  The osculating orbital
elements (for epoch JD 2452490) listed in Table \ref{2plparam} are in Jacobi
coordinates.  In these coordinates, each planet is explicitly assumed to be in
orbit about the center of mass of all bodies interior to its orbit.  As
explained in Lissauer \& Rivera (2001) and Lee \& Peale (2003), Jacobi
coordinates are the most natural system for expressing multiple-planet fits to
radial velocity data.

The uncertainties listed in Table \ref{2plparam} were obtained by performing
1000 additional 2-planet fits to 1000 radial velocity data sets generated using
the bootstrap technique described in Section 15.6 of Press et al. (1992).  Each
velocity data set consisted of 155 entries chosen at random from the 155
entries in the actual velocity data set (Table \ref{velocities}).  Each entry
consists of the observing epoch, the velocity measurement, and the instrumental
uncertainty.  For every choice, all 155 entries were available to be chosen.
This procedure results in generated velocity data sets that contain duplicate
entries.  The fitting algorithm cannot handle such a data set.  Thus, when an
entry is chosen more than once during the generation of a velocity data set,
0.001 day is added to the observing epoch of each duplicate entry.  We then
performed 2-planet fits to each of these 1000 velocity data sets, using the
parameters from the nominal 2-planet fit in the initial guesses.  The 1000 fits
result in ranges in the fitted values of each of the parameters.  The
uncertainties listed in Table \ref{2plparam} are the standard deviations of the
distributions of the parameters.

Using an earlier data set of radial velocity measurements, we had also
performed an analysis to determine uncertainties in the fitted parameters like
the one above and an analysis using randomized residuals (described in Section
\ref{Aliasing}) in which a model is assumed to generate synthetic data sets.
There was no (significant) difference in the resulting uncertainties.  For the
data set used in this work, we use just the bootstrap method to determine the
uncertainties in the fitted parameters since this technique is not constrained
by an assumed model.  It should be noted that it is not a problem to apply the
bootstrap method in determining the uncertainties in the parameters since the
method in which the parameters are determined ($\chi^2$ minimization) does not
depend on the order in which (the square of) the differences between the model
and observed velocities are summed.  That is, the data points are identically
distributed (see pp. 686 and 687 in Press et al. 1992).

The uncertainties determined using the bootstrap method explicitly {\it do not}
account for the correlations which exist among the fitted orbital parameters.
On the other hand, the uncertainties determined from the Levenberg-Marquardt
routine {\it do} account for the correlations.  An example involving $e$ and
$\omega$ illustrates the differences.  In general, from the Levenberg-Marquardt
routine, a small fitted value for $e$ results in a large uncertainty in
$\omega$, and a large value for $e$ results in a small uncertainty in $\omega$.
Thus, if the uncertainty in $e$ is large, the uncertainty in $\omega$ may be
misleading since it depends on the fitted value of $e$ (whereas the actual
value of $e$ could be any value within the range of its uncertainty).  The
bootstrap method is able to explore the full range of possible values of both
$e$ and $\omega$ simultaneously.

It is possible that the bootstrap method can give different uncertainties if we
use different initial guesses.  However, for the GJ~876 system with an assumed
coplanar configuration, we have verified that the fitted parameters are
relatively robust to changes in the initial guesses (Rivera \& Lissauer 2001).
When we explore systems with mutual inclinations, multiple minima appear (see
Section \ref{3plcf}).  Thus, varying the initial guesses while using the
bootstrap method to determine uncertainties becomes more relevant for fits with
mutual inclinations.

In agreement with previous studies, we find that the system corresponding to
the parameters in Table \ref{2plparam} is also deeply in the 2:1 mean motion
resonance and in the apsidal resonance.  We performed 1000 simulations (for
10 years each) based on the nominal fit in Table \ref{2plparam} and the
1000 fits used to determine the uncertainties in the parameters.  The three
critical arguments, $\theta_1=\lambda_c-2\lambda_b+\varpi_c$,
$\theta_2=\lambda_c-2\lambda_b+\varpi_b$, and $\theta_3=\varpi_c-\varpi_b$,
where $\lambda_c$ and $\lambda_b$ are the respective mean longitudes of
companions c and b and $\varpi_c$ and $\varpi_b$ are the corresponding
longitudes of periastron, all librate about 0$^{\circ}$ with the following
small amplitudes: $|\theta_1|_{max}=5.6\,\pm\,1.4^{\circ}$,
$|\theta_2|_{max}=28.8\,\pm\,8.4^{\circ}$, and
$|\theta_3|_{max}=29.0\,\pm\,9.3^{\circ}$.  These amplitudes are smaller than
but consistent with the values from Laughlin et al. (2005).  Note that all
three critical arguments have approximately the same period of libration
($\sim$ 8.8 yr).

The overall long-period ($\sim8.8$ yr) envelope of the model radial velocity in
Figure \ref{2mf=1.00} arises from beating between the periods of the
jovian-mass planets, since the periods are not exactly commensurate. The radial
velocity reversals occur when planets b and c are on opposite sides of the
star, which occurs roughly every 60 days. An inspection of the two components
of the radial velocity due to companions b and c, the sum of these two
velocities, and the mean anomalies of planets b and c shows that the reversals
occur when c is near periastron ($M_c\sim0^{\circ}$) while b is near apastron
($M_b\sim180^{\circ}$). Thus, reversals occur when the (full) orbital velocity
of companion b is near a minimum while that of c is near a maximum.  This
configuration arises because of the resonant state of the system. Also, the
acceleration of the star due to companion c when it is at periastron,
$Gm_c/(a_c(1-e_c))^2=0.0339$ cm\,s$^{-2}$ (simply using the values
from Table \ref{2plparam}), is greater than that due to companion b when it is
at apastron, $Gm_b/(a_b(1+e_b))^2=0.0239$ cm\,s$^{-2}$. In comparison, the mean
acceleration of the star due to planet c is $Gm_c/a_c^2=0.0205$ cm\,s$^{-2}$,
and that due to planet b is $Gm_b/a_b^2=0.0252$ cm\,s$^{-2}$.  Inspection
of the longitudes of periastron for planets b and c also shows that the
vertical position in Figure \ref{2mf=1.00} and amplitude of a velocity reversal
is correlated with the longitude of periastron (of both companions). The
largest velocity reversals occur when the lines of apsides are roughly
perpendicular to the line of sight ($\omega\sim0\mbox{ or }180^{\circ}$), and
these occur near maxima in the full radial velocity of the star (e.g., near
epochs JD 2451060, where $\omega\sim360^{\circ}$, and JD 2452600, where
$\omega\sim180^{\circ}$, in Figure \ref{2mf=1.00}). The period of this vertical
traversal of the reversals coincides with the period of circulation of the
periastron longitudes ($\sim8.8$ yr). Note that in a two-Keplerian model,
reversals are also present, and they appear to move vertically (Marcy et al.
2001). However, in the two-Keplerian case, the vertical displacement results
from the periods not being exactly commensurate. In this case the line joining
the planets when they are on opposite side of the star will deviate farther and
farther from the line of apsides. As a result, the shapes of the reversals in
the two types of models appear different (Laughlin et al. 2005).

As in previous studies, we considered various inclinations of the coplanar
system and generated a series of 2-planet (10+1 parameter) fits.  Figure
\ref{chisq_i2} shows the resulting $\sqrt{\chi_{\nu}^2}$ for the 2-planet fits
versus the inclination ($i$) as triangles.  Note that $\sqrt{\chi_{\nu}^2}$
starts to rise when $i \lesssim 50^{\circ}$.  Laughlin et al. (2005) found that
$\sqrt{\chi_{\nu}^2}$ starts to rise only when $i \lesssim 40^{\circ}$.  The
stricter constraint that we are able to derive is primarily a result of the
improvements to the measurements mentioned in Section \ref{obs}.  Moreover,
although previous studies (Laughlin \& Chambers 2001; Rivera \& Lissauer 2001;
Laughlin et al. 2005) only found a very shallow minimum in
$\sqrt{\chi_{\nu}^2}$ versus $i$, the minimum for our larger, more precise data
set is noticeably deeper.  Newtonian models yield far better fits to the data
than do Keplerian models because there is a clear signal of periapse regression
in the data (Ford 2005).  However, this observed regression rate can be matched
by models with relatively low mass planets b and c (high $\sin{i}$) on low
eccentricity orbits, or higher mass planets on more eccentric orbits.  The
$\sqrt{\chi_{\nu}^2}$ increases only when planetary eccentricities (and masses)
become too large, and the shape of the model velocity curve starts to deviate
significantly from the data.  Note that a large value of $\sqrt{\chi_{\nu}^2}$
for a given small value of $i$ does not immediately imply that a system with
such a value for $i$ is unstable on short timescales ($<$ 10 Myr).  Indeed,
Rivera \& Lissauer (2001) found that a system with $i\,\sim\,11.5^{\circ}$ was
stable for at least 500 Myr.

\section{Residuals to the Two-Planet Fit}
\label{Res-2pl}

We performed a (Lomb) periodogram analysis (Section 13.8 in Press et al. 1992)
to the residuals of the 2-planet $i=90^{\circ}$ fit. The result is displayed in
Figure \ref{periodogram} and shows a maximum peak (the peak with the largest
power) with very significant power, i.e., much higher than typical variations,
at a period of 1.9379 days.  The periodograms of the residuals of all of the
2-planet coplanar fits in which we varied the inclination of the system show a
maximum peak at $\sim$ 1.94 days.  This periodicity is also present as the
largest peak in 995 periodograms of the 1000 residuals of the 2-planet coplanar
fits used to determine the uncertainties in Table \ref{2plparam}.  We performed
2-planet $i=90^{\circ}$ fits to various subsets of the first or last N data
points, and the periodicity is also present as the largest peak in the
periodograms of all of the corresponding residuals; the amount of power at 1.94
days rises with N.  Additionally, the blue points in Figure \ref{folded_res}
directly show the phased residuals of the two-planet fit, folded with a period
of 1.9379 days.  The red points in Figure \ref{folded_res} show the phased
residuals of the 2-planet, $i=50^{\circ}$ coplanar fit.  We carried out a
double-Keplerian fit to subsets of the radial velocities, each of which
contained the data for one observing season with a high cadence of
observations, and computed the periodogram of the residuals.  The sum of all
such periodograms exhibited its tallest peak at a period of 1.94 d, in
agreement with the best period found from our dynamical models. This test shows
that the 1.94 d period cannot be an artifact of the dynamical modeling but
rather is inherent in the data.  These results provide evidence that GJ~876
likely has
a third companion, ``d.'' The second, smaller peak in Figure \ref{periodogram},
at 2.0548 days with power $\sim 26$, is likely an alias, and this issue is
addressed in Section \ref{Aliasing}.  The ratio of the power in the two periods
is 1.3394. In the following two sections, we refer to these two periodicities
by their approximate values of 1.94 and 2.05 days, respectively.

An alternative to the third planet hypothesis is that this periodicity is due
to pulsation of the star itself. For the dM2.5 dwarf GJ~436, Butler et al.
(2004) reported a planet having $m\sin{i} = 21\,M_{\oplus}$ with $P=2.8$ days
and $K=18$ m\,s$^{-1}$. Otherwise, none of the 150 M0-M5 dwarfs on the Keck
planet search survey exhibits any periodicity with a 2-day period. This
suggests that M dwarfs do not naturally pulsate at such a period. Moreover, we
are not aware of any timescale within M dwarfs corresponding to 2 days. The
dynamical and acoustical timescale, analogous to the Solar 5-minute
oscillations, would similarly be of order minutes for M dwarfs. We therefore
rule out acoustic modes as the cause of the 2-day period. The rotation period
of GJ~876 is at least $\sim40$ days, based on its narrow spectral lines and its
low chromospheric emission at Ca II H\&K (Delfosse et al. 1998); we present
photometric evidence of a rotation period of 97 days in Section
\ref{photometry}. Thus, rotational modulation of surface features cannot
explain the 2-day period in the velocities. Similarly, gravity modes and
magnetic buoyant processes seem unlikely to explain the high-Q periodicity that
we detect over the timespan of 8 years in GJ~876. Thus, the 2-day periodicity
in velocity cannot be easily explained by any known property of this normal
M dwarf.

\section{Three-Planet Fits}
\label{3plcf}

Based on the results of the periodogram analysis presented in the previous
section, we performed 3-planet self-consistent fits with the period of the
third planet initially guessed to be about 1.94 days.  These 3-planet fits
involve 13+1 parameters; the 3 new fitted parameters are the mass, semi-major
axis, and mean anomaly of the third planet, and the remaining 10+1 parameters
are the same as in the 2-planet fits described in Section \ref{2plcf}.  Because
of the strong eccentricity damping effects of tides at its distance from the
star, the third planet was assumed to be on a circular orbit.  Note that this
assumption is relaxed later on for some fits.

As in Section \ref{2plcf}, we performed a nominal 3-planet fit to obtain the
best fitted parameters plus 1000 additional 3-planet fits to obtain the
uncertainties in the parameters.  For the nominal fits
$\sqrt{\chi_{\nu}^2}=1.154$ and RMS=4.59 m\,s$^{-1}$ for 3 planets, compared to
$\sqrt{\chi_{\nu}^2}=1.593$ and RMS=6.30 m\,s$^{-1}$ for 2 planets.  Like Table
\ref{2plparam} in Section \ref{2plcf}, Table \ref{3plparam} shows the best
fitted parameters for the 3-planet fit with $i=90^{\circ}$.

Figure \ref{phase0} shows the phased velocity contributions due to each planet.
This figure is analogous to Figure 10 in Marcy et al. (2002), which shows the
triple-{\it Keplerian} orbital fit to the radial velocities for 55~Cancri. The
major difference is that our Figure \ref{phase0} shows a triple-{\it Newtonian}
fit.  Both the data and the model show the interactions between GJ~876's
planets.  However, in generating Figure \ref{phase0} all of the data are folded
into the first orbital period after the first observing epoch, while the models
only show the velocities during that first period (in all three panels, the
velocities shown in the second period are duplicated from the first period).
Since the osculating orbital elements for the outer two planets are varying due
to mutual perturbations, the data should deviate from the model, as clearly
shown for companions b and c.  Since companion d is largely decoupled from the
outer planets (in both the data and the model), the observed velocities more
closely match the model, and the deviations shown are primarily due to the
residual velocities.  The decoupling is a consequence of the large ratio of the
orbital periods for planets c and d ($>$ 15).

The parameters for the two previously known outer planets are not significantly
affected by fitting for the parameters for all three planets.  However, all of
the uncertainties of these parameters are reduced.  Thus, the addition of the
third planet does not have as significant an effect on the planetary parameters
as the effect that the addition of companion c had on changing the parameters
of companion b.  This result isn't surprising, given the very low mass and very
different orbital period of planet d.

These results have led us to the likely interpretation of a third companion to
GJ~876 with a minimum mass of $m_d\sin{i_d} \sim 6\,M_{\oplus}$ and a period of
about 2 days.  Although this planet is the lowest mass extrasolar planet yet
discovered around a main sequence star, even lower mass planets have been found
around the millisecond pulsar PSR B1247+1221 (Wolszczan \& Frail 1992; Konacki
\& Wolszczan 2003).

\begin{figure}
\includegraphics[angle=-90,scale=0.61]{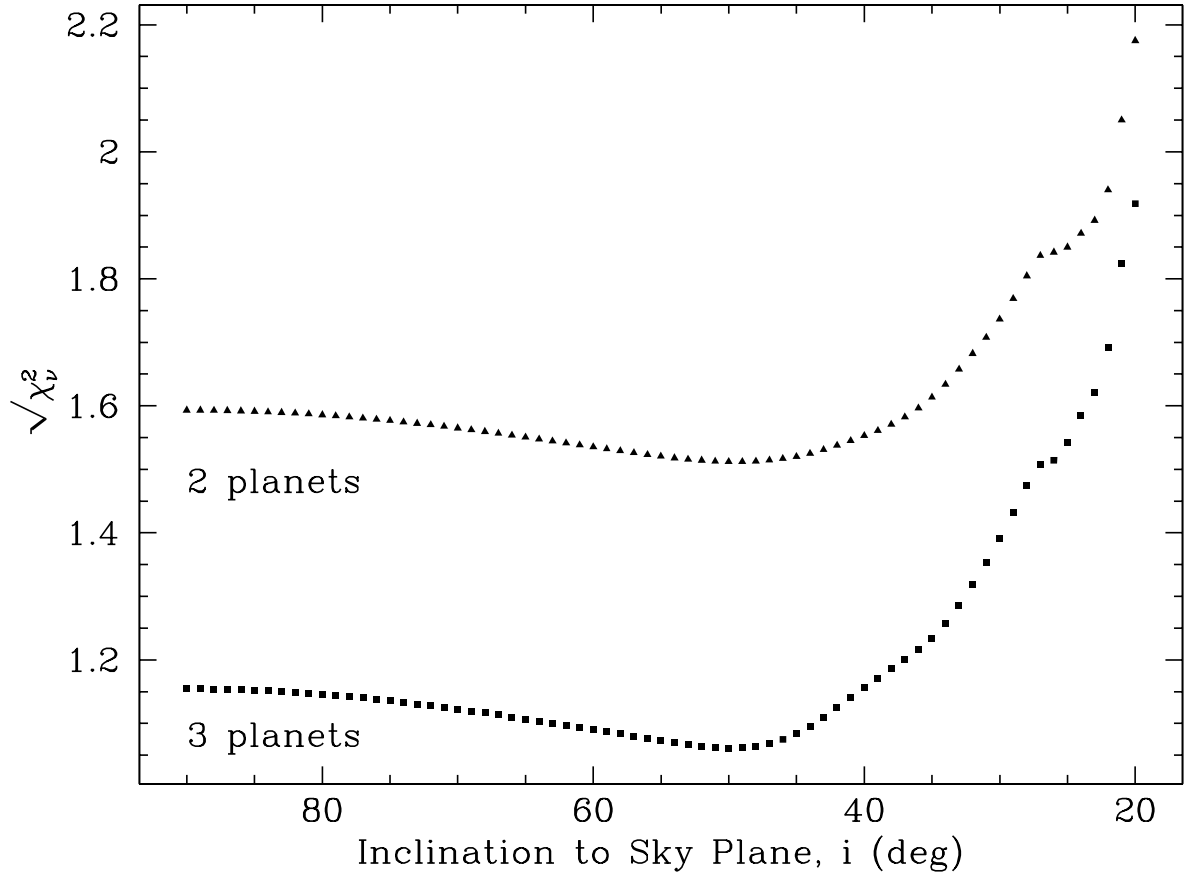}
\caption{
$\sqrt{\chi_{\nu}^2}$ values obtained from three-body (two-planet, triangles)
and four-body (three-planet, squares) fits to the actual GJ~876 radial velocity
data as a function of the inclination of the (assumed coplanar) system to the
plane of the sky, $i$.
}
\label{chisq_i2}
\end{figure}

We also generated a series of 3-planet (13+1 parameter) fits in which we varied
the inclination of the coplanar system.  Figure \ref{chisq_i2} shows
$\sqrt{\chi_{\nu}^2}$ for the 3-planet fits versus the inclination as squares.
The global minimum in $\sqrt{\chi_{\nu}^2}$ (1.061 with an RMS of 4.23
m\,s$^{-1}$) occurs at $i=50^{\circ}$, almost precisely the location of the
minimum for the two-planet fits.
As for the two-planet fits, the $\sqrt{\chi_{\nu}^2}$ starts to rise when
$i < 50^{\circ}$.  Unlike two-planet fits performed on previous data sets, the
minimum at $i=50^{\circ}$ is significant.  Using the bootstrap method as in
Press et al. (1992), we generated 100 velocity data sets and performed a series
of 71 fits ($i=90^{\circ}, i=89^{\circ}, ..., i=20^{\circ}$) to each set in
which we varied the inclination of the system.  This results in 100 curves
which are analogous to the lower curve in Figure \ref{chisq_i2}.  Ninety-eight
of these curves have a minimum in $\sqrt{\chi_{\nu}^2}$ which occurs at
$i=45$\,--\,58$^{\circ}$.  Seventy-nine have a minimum at
$i=48$\,--\,52$^{\circ}$.  Considering all 100 curves, the mean value (and
standard deviation) of the location of the minimum in $\sqrt{\chi_{\nu}^2}$ is
$i=50.2\,\pm\,3.1^{\circ}$, and the median value is 50$^{\circ}$.  The mean
value (and standard deviation) of the drop in $\sqrt{\chi_{\nu}^2}$ from the
value at $i=90^{\circ}$ to the value at the minimum is 0.094\,$\pm$\,0.036,
and the median is 0.097.  The mean value (and standard deviation) of the drop
in $\sqrt{\chi_{\nu}^2}$ from the value at $i=90^{\circ}$ to the value at
$i=50^{\circ}$ is 0.091\,$\pm$\,0.036, and the median is 0.095.  These values
are fully consistent with the drop observed for the actual data.  Figure
\ref{7100fits} shows the entire set of results (small points) along with the
results from fitting the real data (squares).  Figure \ref{3inc50} shows the
model radial velocity generated from the $i=50^{\circ}$ 3-planet fit to the
actual data, along with the actual observed velocities; the residuals are shown
in the lower portion.  Note that the residuals in Figure \ref{3inc50} are shown
on the same scale as in Figure \ref{2mf=1.00}; the dispersion is clearly
smaller in the 3-planet fit.  For completeness, Figure \ref{zoomin} overlays
the two model curves from Figures \ref{2mf=1.00} and \ref{3inc50} and the data
near the epoch JD 2452060; this illustrates the effect that the third planet
has on the model.  Table \ref{3plparami=50} lists the best fitted orbital
parameters for $i=50^{\circ}$.  As in Tables \ref{2plparam} and \ref{3plparam},
the uncertainties listed in Table \ref{3plparami=50} are based on 1000 fits
with $i=50^{\circ}$.  The top panel of Figure \ref{3periodograms} shows the
periodogram of the residuals for this fit.  There are no clearly significantly
strong peaks (but see Section \ref{dis}).

We analyzed the significance of the minimum at $i=50^{\circ}$ by
performing a limited set of fits in which the orbits of planets b and c have a
mutual inclination.  An exhaustive search of the entire 20+1 parameter space
delimited by the masses and six orbital parameters of each of the three
planets, less one representing rotation of the entire system about the line of
sight, is beyond the scope of this paper.  One complication arises from the
appearance of a significant number of multiple minima.  For example, Rivera \&
Lissauer (2001) found several satisfactory two-planet fits (with similar values
of $\sqrt{\chi_{\nu}^2}$) which had significantly different fitted orbital
parameters.  We did, however, fit parameters for two sets of non-coplanar
planetary configurations.  In both cases, the planetary orbital planes at epoch
were fixed such that the tiny inner planet and one of the giant planets were
coplanar with orbit normal inclined at epoch by 50$^{\circ}$ from the line of
sight, and the other giant planet's orbit normal was inclined by a
pre-determined amount from the line of sight with the same node as that of the
other two planets, so that the mutual inclination was equal to the difference
in inclination to the line of sight.  Other initial parameters for the fitting
were taken from the fit given in Table \ref{3plparami=50}, with $m\sin{i}$,
rather than $m$, conserved for the non-coplanar planet.  In one set, the inner
and middle planets had $i=50^{\circ}$ and the outer planet's inclination
varied.  In the other set of fits, the inner and outer planets had
$i=50^{\circ}$ and the middle planet's inclination differed.  The
$\sqrt{\chi_{\nu}^2}$ of these fits are shown in Figure \ref{chi_ibc}.  Since
only a small amount of the parameter space was explored, these preliminary
results are only sufficient to draw two tentative conclusions: 1) since
$\sqrt{\chi_{\nu}^2}$ rises more rapidly as the inclination of b is varied, the
minimum in $\sqrt{\chi_{\nu}^2}$ appears primarily to constrain the inclination
of companion b, and 2) the mutual inclination between the outer two planets is
likely small.  Note that since the nodes were all the same in these fits,
varying the mutual inclination changes the mass ratio of the planets; in
contrast, varying the nodes can produce configurations with large mutual
inclinations but similar mass ratios to those estimated assuming coplanar systems.

As for the two-planet case, the jovian-mass planets are deeply locked in the
resonant state in three-planet simulations.  For $i=90^{\circ}$,
$|\theta_1|_{max}=5.9\,\pm\,1.1^{\circ}$,
$|\theta_2|_{max}=30.2\,\pm\,6.1^{\circ}$, and
$|\theta_3|_{max}=30.1\,\pm\,6.6^{\circ}$.  Note that for the 3-planet
simulations, the uncertainties of the amplitudes of the critical angles are
reduced, as are the uncertainties in the parameters in Table \ref{3plparam}.
For $i=50^{\circ}$,
$|\theta_1|_{max}=5.4\,\pm\,0.9^{\circ}$,
$|\theta_2|_{max}=19.5\,\pm\,3.8^{\circ}$, and
$|\theta_3|_{max}=19.4\,\pm\,4.3^{\circ}$.  As in Laughlin et al. (2005), we
find a general trend in which the amplitudes of the critical arguments decrease
as $i$ decreases.

We attempted to determine a dynamical upper limit to the mass of planet d.  For
a coplanar system with $i=50^{\circ}$, the fitted $m_d$ is $7.53\,M_{\oplus}$.
However, we find that the introduction of an inclination of planet d's orbit to
the initial orbital plane of planets b and c has little effect on
$\sqrt{\chi_{\nu}^2}$.  We performed a set of 3-planet fits in which we kept
the outer two planets in the same plane with $i=50^{\circ}$, but varied the
inclination of companion d through values $< 50^{\circ}$.  All three nodes were
kept aligned.  We find that $\sqrt{\chi_{\nu}^2}$ does not deviate
significantly above 1.061 until $i_d<3^{\circ}$.  We then used the Mercury
integration package (Chambers 1999), modified as in Lissauer \& Rivera (2001)
to simulate the general relativistic precession of the periastra, to perform
simulations up to 1 Myr based on these fits, and find that the system is stable
for $i_d\ge3^{\circ}$.  The fitted mass for planet d for the most inclined
stable system is $\sim 103\,M_{\oplus}$.  This result indicates that the orbit
normal of planet d may lie at least as close as $3^{\circ}$ to the line of
sight.  The orbit normal could point toward or away from us.  This defines a
double cone which occupies a solid angle of 0.0172 steradians, or about 0.137\%
of $4\pi$ steradians.  Even by restricting the parameter space by fixing
$i_b=i_c=50^{\circ}$, stability considerations can only exclude configurations
with the orbit normal of companion d in this small solid angle.  Thus, these
stability considerations cannot presently provide a very meaningful upper bound
on the mass of companion d.

With only $m\sin{i}$ determined here for the new planet, its actual mass and
the value of $i$ remain essentially unconstrained by dynamical considerations.
Nearly pole-on orientations of the orbital plane cannot be ruled out. However,
for randomly oriented orbital planes, the probability that the inclination is
$i$ or lower (more face-on) is given by $P(i)=1-\cos{i}$. Thus, for example,
the probability that $\sin{i}<0.5$ is 13\%. Hence, it is {\it a priori}
unlikely that $m_d>2m_d\sin{i_d}$. Moreover, GJ~876 is the only M dwarf for
which such intense Doppler observations have been made (due to the interest in
the outer two planets). The population of M dwarfs from which this low
$m\sin{i}$ was found is only one, namely GJ~876 itself. In contrast, more than
150 M stars have been observed with adequate precision to detect planets of
Saturn mass or greater in two day orbits, and most of these have been observed
with adequate precision to detect a Neptune mass planet so close in. Yet apart
from GJ~876, the only M star known to possess a planet is GJ~436, which has a
companion with $m\sin{i}=21\,M_{\oplus}$ on a 2.644-day orbit (Butler et al.
2004). Therefore, the low $m\sin{i}$ of GJ~876 d was likely not drawn from some
large reservoir for which a few nearly pole-on orbits might be expected. We
conclude that the true mass of the new planet is likely to be
$\lesssim10\,M_{\oplus}$.

We also performed analyses to place limits on the eccentricity of planet d. Two
series of one-planet fits to the velocity residuals of both the $i=90^{\circ}$
and $i=50^{\circ}$ two-planet fits (Figure \ref{folded_res}) suggest that the
eccentricity of the third companion could be as high as $\sim0.22$. For the
initial guesses, we used the best-fitted mass, period, and mean anomaly for
companion d from the three-planet self-consistent fit (from Table
\ref{3plparam} for $i=90^{\circ}$ and from Table \ref{3plparami=50} for
$i=50^{\circ}$), but varied the initial guessed eccentricity and argument of
periastron, and fitted for the eccentricity, argument of periastron, and mean
anomaly. Figure \ref{1plfit} shows the resulting phased velocities for
$i=90^{\circ}$. Additionally, for both $i=90^{\circ}$ and $i=50^{\circ}$ we
performed a few fits including all three planets and using an initial guessed
eccentricity for the third planet $e_d\sim0.22$. The largest fitted value is
$\sim0.28$ (for each value of $i$); this represents our best estimate for an
upper limit on the eccentricity of companion d. Based on each of the best fit
parameters in Tables \ref{3plparam} and \ref{3plparami=50}, dynamical
integrations of the system with the inner planet initially on a circular orbit
show that the forced eccentricity of companion d is only $\sim0.0018$ for
$i=90^{\circ}$ and $\sim0.0036$ for $i=50^{\circ}$. The tidal circularization
timescale for a planet of mass $m_{\rm pl}$ and radius $R_{\rm pl}$ in an orbit
with semi-major axis $a$ about a star of mass $M_{\star}$ is
\begin{equation}
\tau=\frac{4}{63}Q\left(\frac{a^3}{GM_{\star}}\right)^{1/2}\left(\frac{m_{\rm pl}}{M_{\star}}\right)\left(\frac{a}{R_{\rm pl}}\right)^5
\end{equation}
(Goldreich \& Soter 1966, Rasio et al. 1996). For GJ~876 d, for $i=90^{\circ}$,
$a=0.021$ AU, $m_{\rm pl}=5.9\,M_{\oplus}$ and $R_{\rm pl}=1.619\,R_{\oplus}$
(see Section \ref{transits}) this timescale is $<10^5$ yr if companion d has a
dissipation factor, $Q$, similar to that of Earth ($\sim10$). If the $Q$ of
companion d is similar to the estimated $Q$ values for the outer planets in the
Solar System ($10^4$\,--\,$10^5$), then the timescale for eccentricity damping
would be 40\,--\,400 Myr, which is less than the estimated 1-10 Gyr lifetime of
the star (Marcy et al. 1998). These arguments and results indicate that the
eccentricity of companion d is fully consistent with 0.

We addressed the issue of stellar jitter by performing a few 3-planet,
$i=50^{\circ}$, fits in which we folded an assumed value of stellar jitter in
quadrature into the instrumental uncertainties (listed in Table
\ref{velocities}). The most relevant quantity in these fits is the
$\sqrt{\chi_{\nu}^2}$. Although we only fit for 13+1 parameters at a time, by
varying the inclination of the system, we effectively allowed a $15^{\rm th}$
free parameter. To account for this extra parameter, the formal
$\sqrt{\chi_{\nu}^2}$ must be adjusted upwards by a factor of
$\sqrt{141/140}\approx1.0036$. (Note that the 141 and 140 are the number of
observations, 155, minus the number of fitted parameters, 13+1 and 14+1,
respectively.)  Accounting for this factor, and folding in an
assumed jitter of 0, 0.5, 1.0, 1.5, and 2.0 m\,s$^{-1}$, the
$\sqrt{\chi_{\nu}^2}$ are 1.065, 1.057, 1.034, 0.9996, and 0.956 respectively.
These results indicate that the actual stellar jitter of GJ~876 is likely to be
small ($\lesssim1.5$ m\,s$^{-1}$), as it is unlikely for the
$\sqrt{\chi_{\nu}^2}$ to be substantially less than unity for a data set as
large as the one used in this paper. Note that the period of companion d is the
same to better than 1 part in $10^5$ for all five values of the assumed stellar
jitter.

\section{Aliasing: What Is the Period of the Third Companion?}
\label{Aliasing}

The periodogram presented in Section \ref{Res-2pl} shows significant power at
both 1.94 days and 2.05 days.  Using $\sim$ 2.05 days as an initial guess for
the period of the third planet, we performed a 3-planet fit to the observed
radial velocities.  The resulting fitted period for the third planet is
2.0546 days.  More importantly, this fit is not vastly worse than the fit with
the period of the third planet initially guessed to be 1.94 days
($\sqrt{\chi_{\nu}^2}=1.154$ and RMS\,=\,4.59 m\,s$^{-1}$ for 1.94 days, and
$\sqrt{\chi_{\nu}^2}=1.290$ and RMS\,=\,5.08 m\,s$^{-1}$ for 2.05 days,
compared with $\sqrt{\chi_{\nu}^2}=1.593$ and RMS\,=\,6.30 m\,s$^{-1}$ for the
corresponding 2-planet model).  This result prompted a series of tests which
together strongly support the hypothesis that the 1.94-day period is the
correct one (and the 2.05-day period is an alias), as follows.

We first examined the periodograms of the residuals of the three-planet fits
for each period (the lower two panels of Figure \ref{3periodograms}), and we
detected no peaks with very significant power at any period (there is
moderate power near 9 days and at other periods --- see Section \ref{dis}). The
detection of a significant peak at 1.94 days in the residuals of the
three-planet fit with $P_d=2.05$ days would have been a clear indication that
the 1.94-day period is the true period, because the introduction of a third
planet with the wrong period should not (fully) account for the true
periodicity. This simple test thus implies that one of the near two-day periods
is an alias, but it does not indicate which period is the alias.

We then analyzed various mock velocity data sets to determine whether or not
both near 2-day periodicities could result purely from the spacing of the
observations and to determine the relative robustness of the two short-period
planet solutions.  We generated the mock velocity data sets for this study by
randomizing residuals, as follows: The difference between the observed and
modeled velocities results in a residual velocity at each observing epoch.  We
indexed the 155 residuals.  At each observing epoch, we chose a random integer
from 1 to 155 inclusive, and added the corresponding residual to the model
velocity at that epoch.  

\begin{figure}
\includegraphics[angle=-90,scale=0.61]{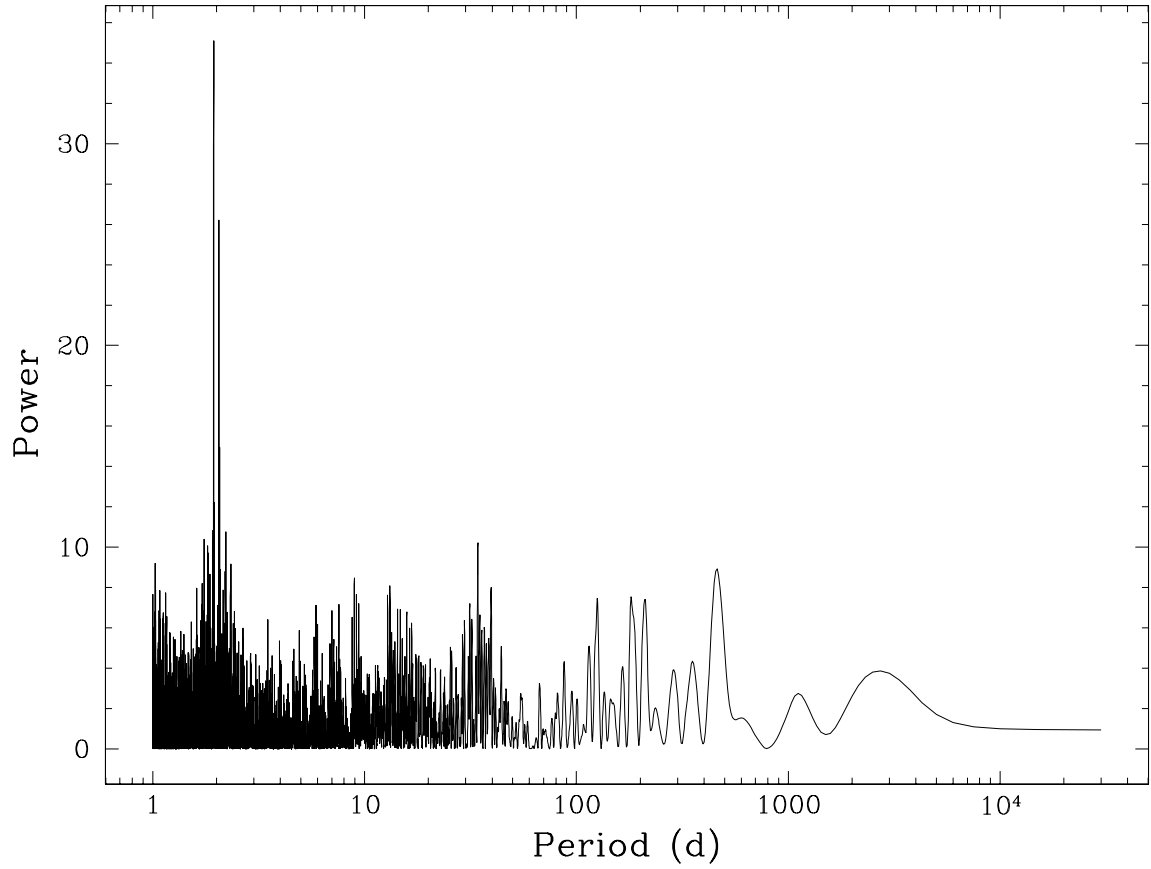}
\caption{
Periodogram of Residuals to the nominal two-planet, $i=90^{\circ}$, coplanar
fit, presented in Table \ref{2plparam} and Figure \ref{2mf=1.00}.
Note the strong power at 1.938 and 2.055 days.
}
\label{periodogram}
\end{figure}

\begin{figure}
\includegraphics[angle=-90,scale=0.61]{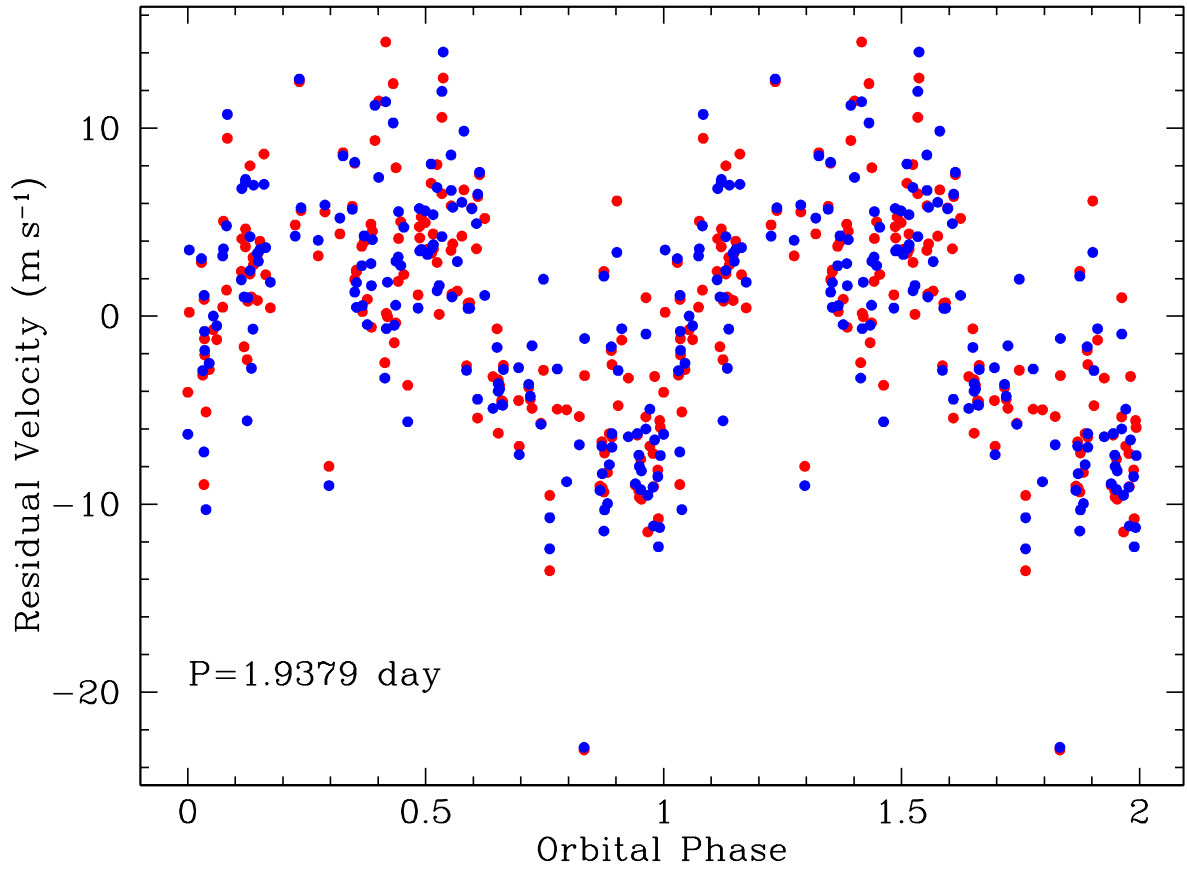}
\caption{
Blue: residuals to the nominal two-planet, $i=90^{\circ}$, coplanar fit.  Red:
residuals to the two-planet, $i=50^{\circ}$, coplanar fit.  These residuals
have been folded at the period of tallest peak shown
in Figure \ref{periodogram}, with the epoch of the first observation defining
zero orbital phase. All points are plotted twice (as two cycles) to improve
the visibility of the pattern.
}
\label{folded_res}
\end{figure}

\begin{figure}
\includegraphics[scale=0.565]{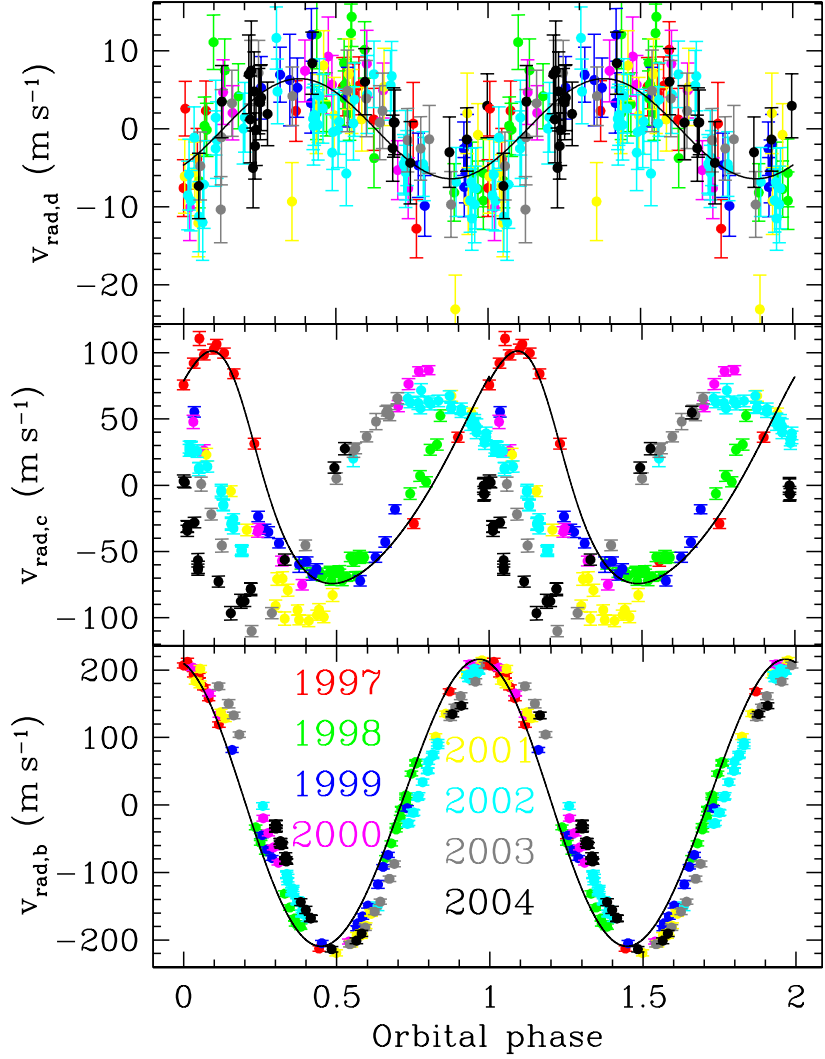}
\caption{
Triple-Newtonian orbital fit to the radial velocity observations for GJ~876.
The observed and model velocities for each planet are shown separately by
subtracting the effects of the other two planets. The panels show the
velocities due to companions d (top), c (middle), and b (bottom).  The curves
show model velocities for the first orbital period beginning with the epoch of
the first observation.  The data were folded at the appropriate periodicity
given by the fit in Table \ref{3plparam}.  Note the differences in scale in the
three panels.  The deviations shown for companions c and b clearly demonstrate
that their orbital elements have been evolving over the timespan of the
observations.  The colored numbers in the bottom panel indicate which points
correspond to which observing season.  Note that the points taken in 1997 most
closely follow the curves in the bottom two panels, as expected.
}
\label{phase0}
\end{figure}

\begin{figure}
\includegraphics[angle=-90,scale=0.61]{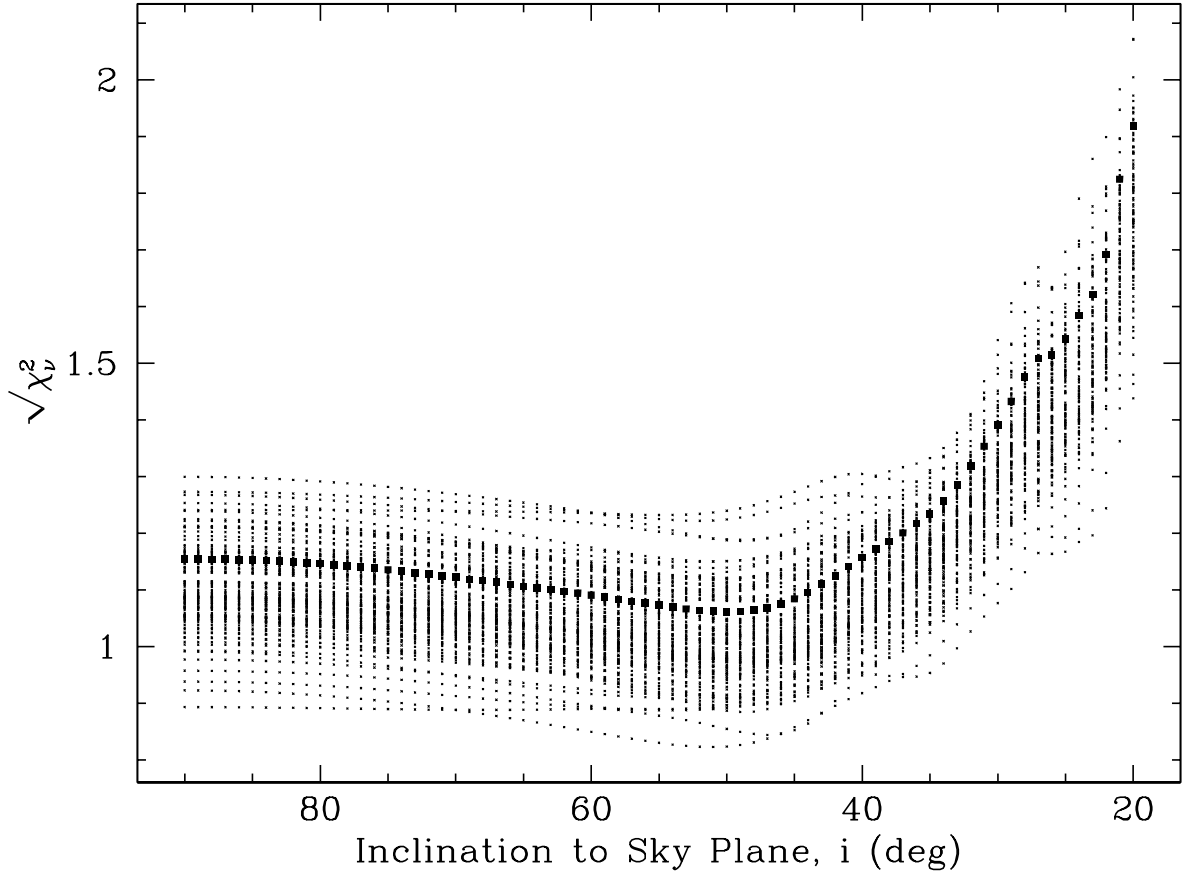}
\caption{
One hundred sets of results showing $\sqrt{\chi_{\nu}^2}$ versus the
inclination to the plane of the sky, $i$, for 3-planet models to 100 generated
radial velocity data sets (small points).  The squares show the result of
fitting the actual data.  Note that most of the results for the generated
velocity data sets have a minimum near $i=50^{\circ}$, as for the result for
the real velocity data set.
}
\label{7100fits}
\end{figure}

\begin{figure}
\includegraphics[angle=-90,scale=0.61]{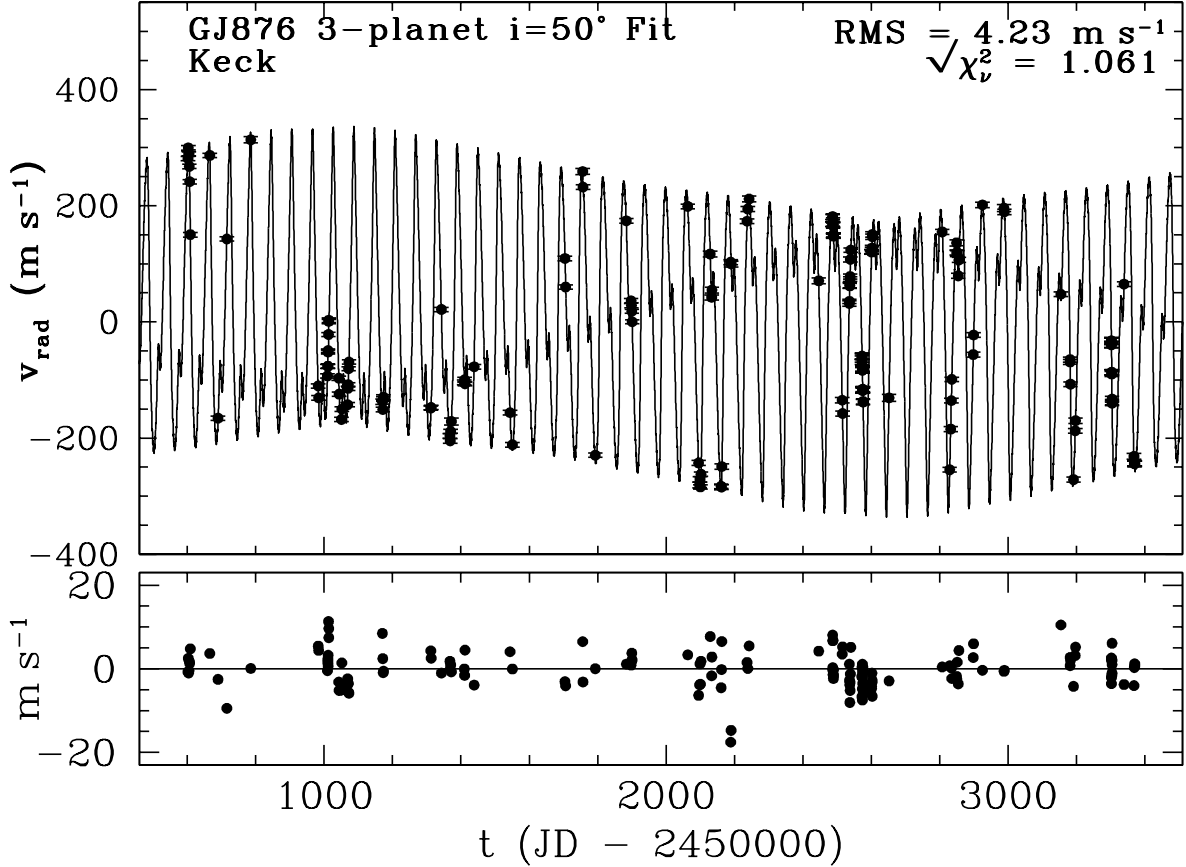}
\caption{
Top: Model radial velocity (solid line) generated from the self-consistent,
coplanar, $i=50^{\circ}$, four-body (three-planet) fit to the observed radial
velocities for GJ~876.  The observed velocities (from Table \ref{velocities})
are shown as small solid points with vertical error bars corresponding
to the instrumental uncertainties.  Bottom: Residuals to the orbital fit.  Note
that the $\sqrt{\chi_{\nu}^2}$ is 1.06, much lower than for the 2-planet model
($\sqrt{\chi_{\nu}^2}=1.59$) and somewhat lower than the 3-planet model with
$i=90^{\circ}$.
}
\label{3inc50}
\end{figure}

\begin{figure}
\includegraphics[angle=-90,scale=0.61]{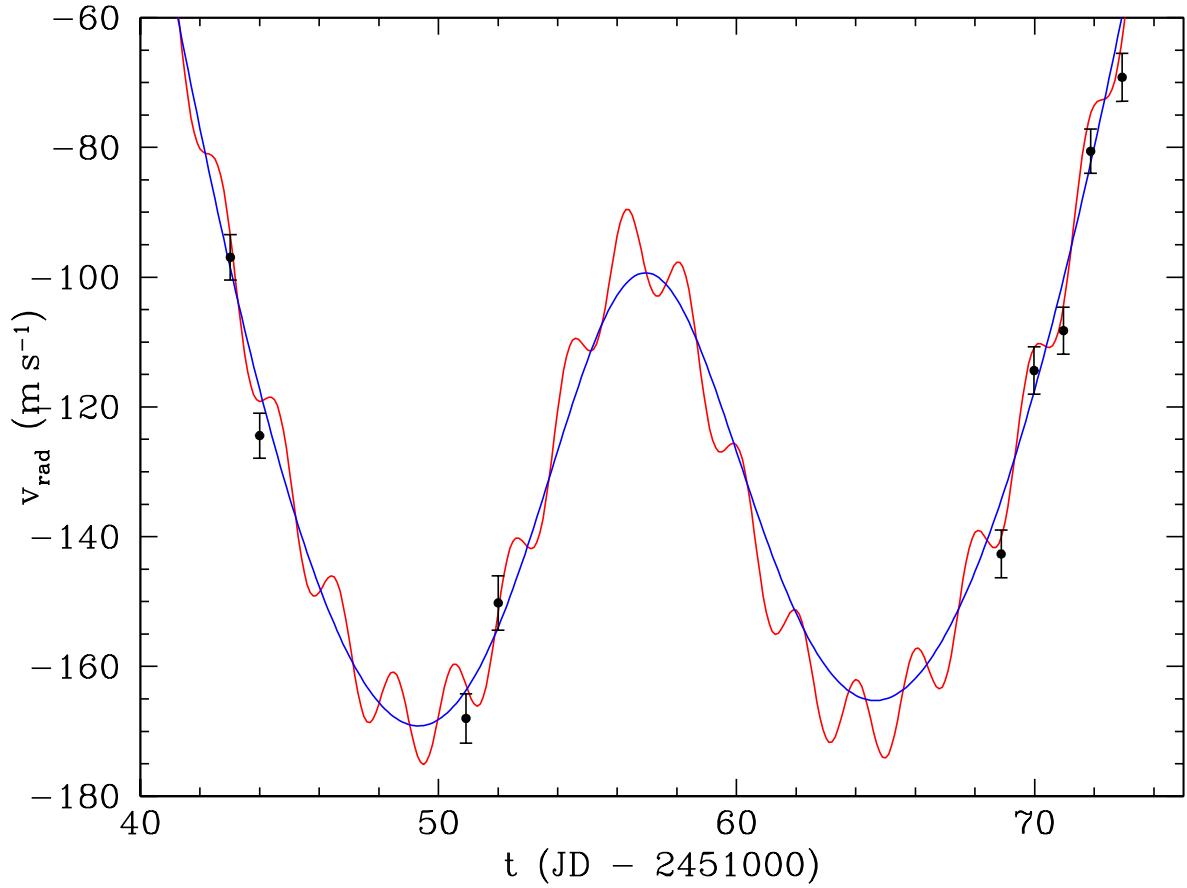}
\caption{
Zoom in near JD 2452060 on the model radial velocities from the two-planet
$i=90^{\circ}$ model shown in Figure \ref{2mf=1.00} (blue curve) and the
three-planet $i=50^{\circ}$ model shown in Figure \ref{3inc50} (red curve) with
the data overplotted.
}
\label{zoomin}
\end{figure}

\begin{figure}
\includegraphics[scale=0.69]{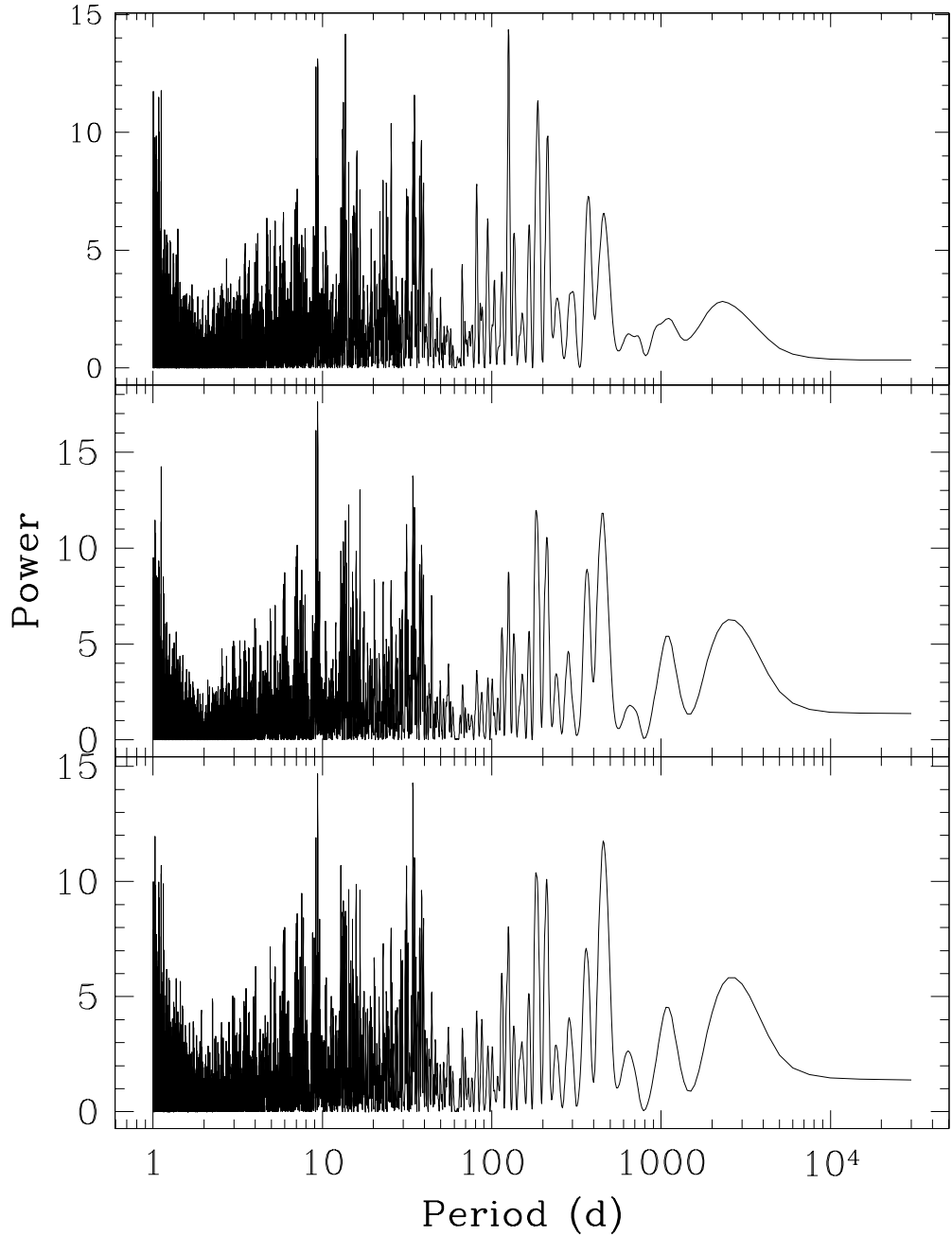}
\caption{
Top: Periodogram of residuals to the $i=50^{\circ}$ 3-planet self-consistent
fit (with $P_d=1.9379$ days).
Middle: Periodogram of Residuals to the nominal, $i=90^{\circ}$, 3-planet,
self-consistent fit with $P_d=1.9379$ days.
Bottom: Periodogram of Residuals to the nominal, $i=90^{\circ}$, 3-planet,
self-consistent fit with $P_d=2.0548$ days.
}
\label{3periodograms}
\end{figure}

\begin{figure}
\includegraphics[angle=-90,scale=0.61]{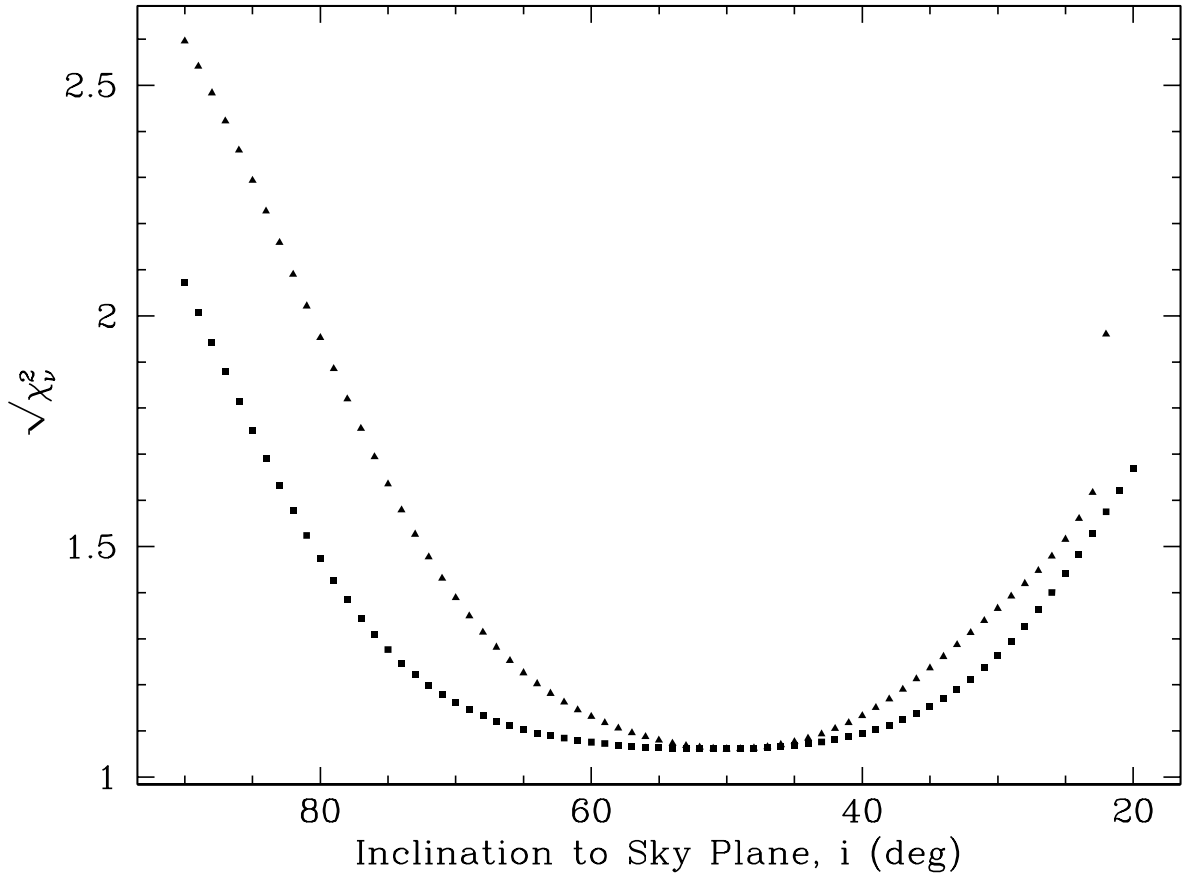}
\caption{
$\sqrt{\chi_{\nu}^2}$ versus the inclination to the plane of the sky for
3-planet configurations in the GJ~876 system in which the inclination of one of
the outer planets is varied while the inclination of the remaining two planets
is $i=50^{\circ}$.  Triangles show the effect of varying the inclination of
planet b, while squares show the result of varying the inclination of companion
c.  All three nodes were aligned.
}
\label{chi_ibc}
\end{figure}

\begin{figure}
\includegraphics[angle=-90,scale=0.61]{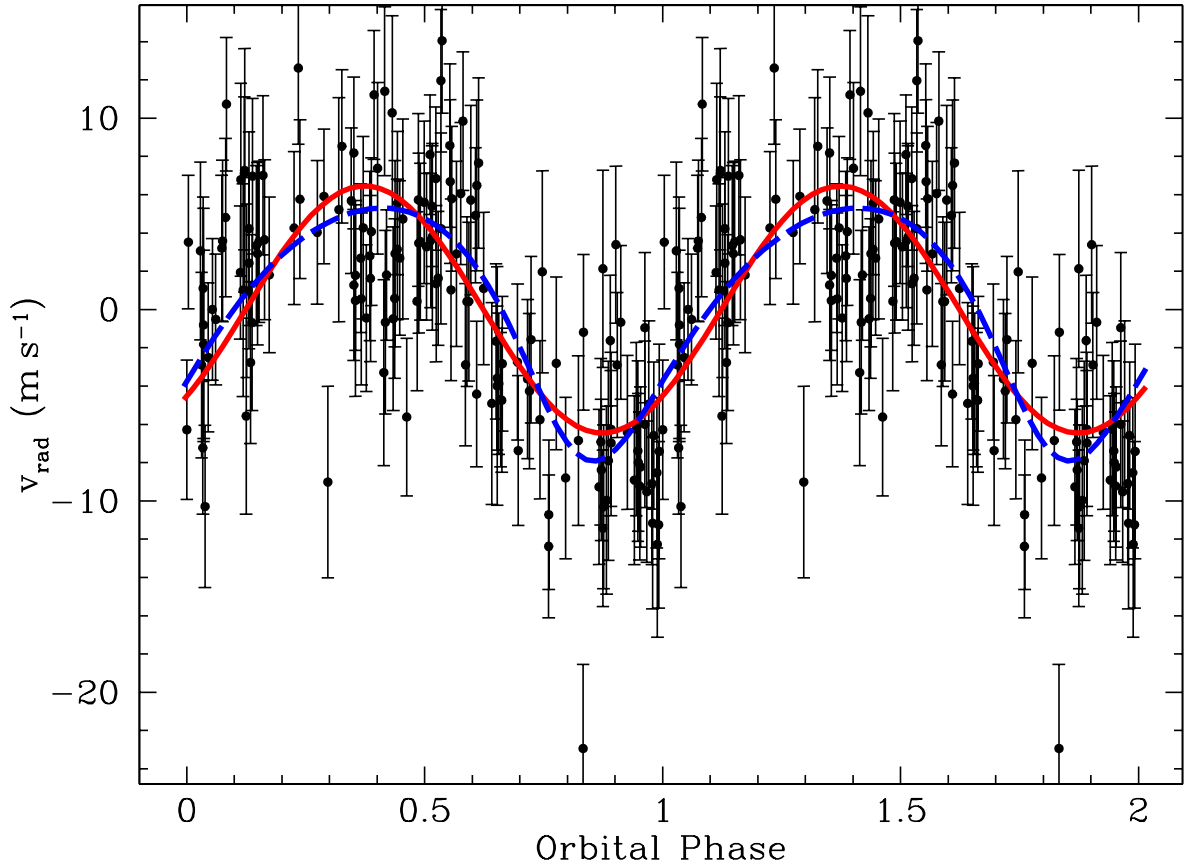}
\caption{
Model radial velocities generated from one-planet fits to the residuals of the
$i=90^{\circ}$ two-planet fit to the observed radial velocities for GJ~876.
The solid red line is for a fit with $e_d=0$, and the dashed blue line is for a fit with
$e_d=0.22$.  The residual velocities are phased with a period of 1.9379 days
and are shown as small solid points with vertical error bars corresponding to
the instrumental uncertainties.
}
\label{1plfit}
\end{figure}

\begin{figure}
\includegraphics[scale=0.715]{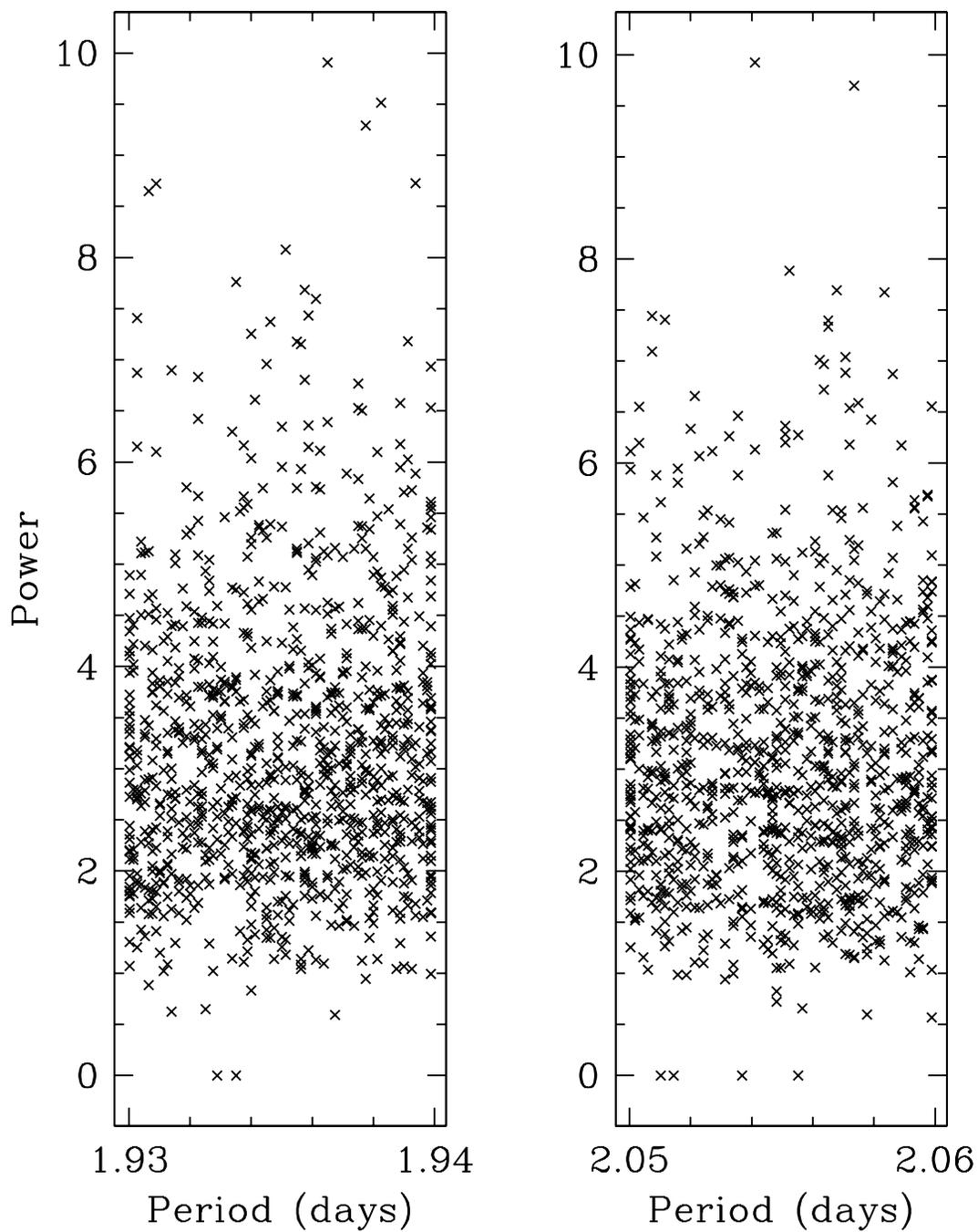}
\caption{
Maximum power at 1.94 days (left) and 2.05 days (right) in 1000 periodograms of
residuals to two-planet fits to 1000 synthetic velocity data sets generated
from a two-planet model.
}
\label{plot_power}
\end{figure}

One issue to address is whether the third periodicity (for both periods) is an
artifact of the observing frequency. We generated 1000 mock velocities by
randomizing the residuals of the two-planet model and performed two-planet fits
to these velocities. If the third periodicity is purely due to the observing
frequency, then the periodograms of the residuals to these two-planet fits
should show significant power at the third periodicity. Figure \ref{plot_power}
shows the maximum power at the two periods in each of the 1000 periodograms. In
not one case out of 1000 did we observe a periodogram which resembled the
periodogram presented in Figure \ref{periodogram}. Not one peak was found at
either period that was as significant as the ones in the periodogram in Figure
\ref{periodogram}. In fact, for the 1.94-day period, the mean (and standard
deviation) of the maximum power for the 1000 periodograms is $3.2\,\pm\,1.4$.
For the 2.05-day period, this is $3.1\,\pm\,1.3$. The most significant peaks at
either periodicity had a power of $\sim10$, about 38\% of the power in the
second highest peak in Figure \ref{periodogram}. In Figure \ref{periodogram},
the observed maximum peak at 1.94 days ($\sim35$) is $>22$ standard deviations
above the mean value of the maximum peaks determined in Figure
\ref{plot_power}. At 2.05 days, the observed power ($\sim26$) is $>17$ standard
deviations above the mean value of the maximum peaks determined in Figure
\ref{plot_power}. This strongly indicates that (at least) one of the
periodicities must be real.

\begin{figure}
\includegraphics[angle=-90,scale=0.61]{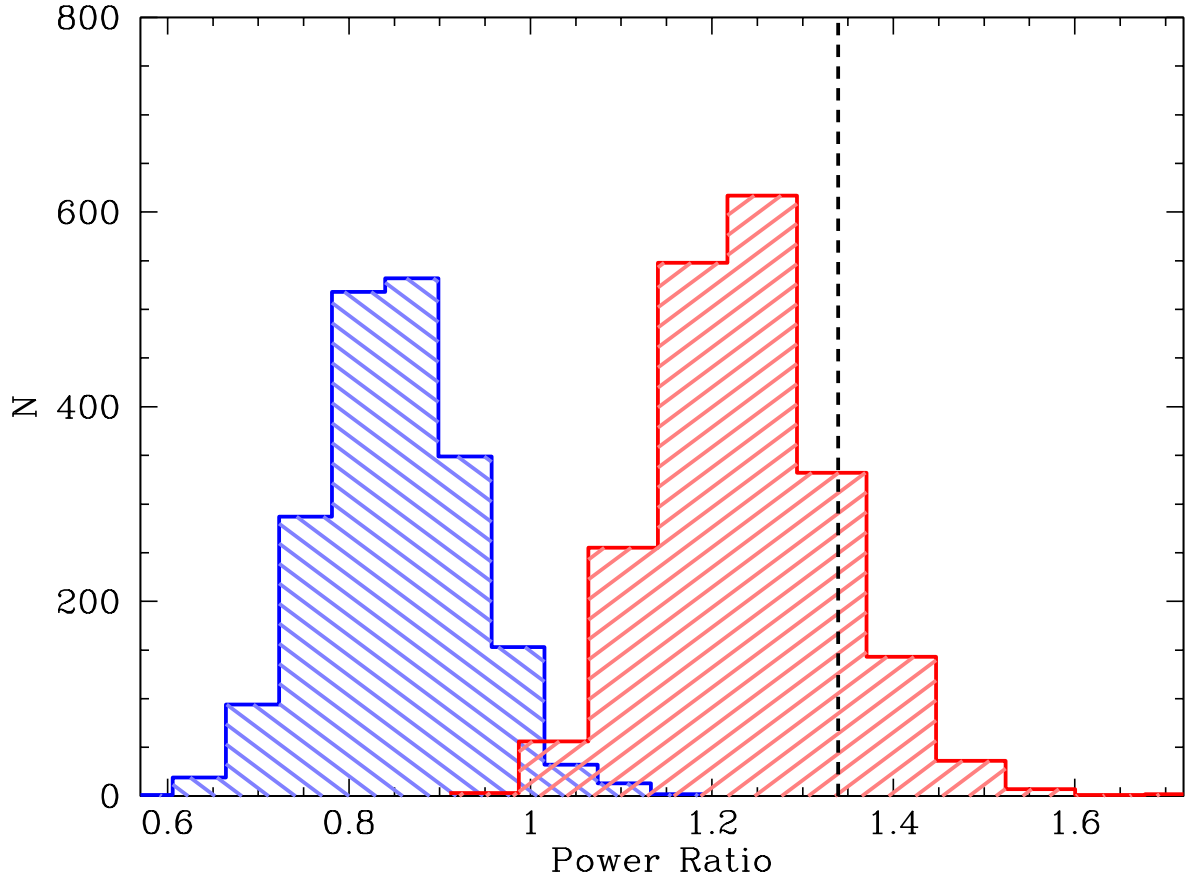}
\caption{
Histograms of the ratio of the power at 1.94 days to the power at 2.05 days for
4000 periodograms of residuals to two-planet fits to mock velocities generated
from 3-planet models.  Red shows the results when the third planet has a period
of 1.94 days.  Blue shows the results when the third planet has a period of
2.05 days.  The dashed line at 1.3394 indicates the ratio observed in the
periodogram of the residuals of the $i=90^{\circ}$ two-planet fit to the actual
data.
}
\label{power_ratio_hist}
\end{figure}

We performed similar experiments in which we generated 4000 sets of mock
velocities based on the three-planet model.  Using randomizing residuals, two
thousand of the sets
were generated based on the model in which the third planet had a period of
1.94 days.  The remaining 2000 sets were generated in an analogous manner but
based upon the model in which the third planet had a period of 2.05 days.  We
then performed two-planet fits to all 4000 velocity sets.  Then, we examined
the periodograms of the residuals of these fits to see if we could generate
results similar to the one in Section \ref{Res-2pl}.  Figure
\ref{power_ratio_hist} shows histograms of the ratio of the power at 1.94 days
to the power at 2.05 days.  Red is for the models with the third planet at 1.94
days, and blue is for the models with the third planet at 2.05 days.  With a
model in which the third planet had a period of 1.94 days, 1996 periodograms
have a maximum peak at 1.94 days, and 278 have a ratio in the power at 1.94
days to the power at 2.05 days exceeding 1.3394 (478 have this ratio exceeding
1.3).  Thus, the model with $P_d=1.94$ days can yield periodograms which
resemble the result when a two-planet fit is performed on the actual data. With
a model in which the third planet had a period of 2.05 days, 79 periodograms
have a maximum peak at 1.94 days, and 0 have a ratio in the power at 1.94 days
to the power at 2.05 days exceeding 1.3394.  Thus, the model with $P_d=2.05$
days {\em never} resulted in a periodogram which resembles the result when a
two-planet fit is performed on the actual data.  This is very strong evidence
that the 1.94-day period is the true period of the third companion.

Additional results also indicate that the 1.94-day period is the ``better''
one.  Briefly, the $\sqrt{\chi_{\nu}^2}$ and RMS are smaller, and there is more
power at 1.94 days in the periodogram of the residuals of the two-planet fit.
For the three-planet fits, the osculating radial velocity amplitude of the
star, $K$, due to the third planet is $\sim$ 40\% larger than the RMS of the
fit with $P_d=1.94$ days, while this $K$ is only 5\% larger than the RMS for
the fit with $P_d=2.05$ days.

\section{Photometric Variability in GJ~876}
\label{photometry}

Very little is known about the photometric variability of GJ~876 on rotational
and magnetic cycle timescales. Weis (1994) acquired 38 Johnson $V$ and Kron
$RI$ measurements at Kitt Peak National Observatory over an 11 year period. He
observed variability of a couple percent or so and suspected a possible
periodicity of 2.9 years. Based on these findings, Kazarovets \& Samus (1997)
assigned GJ~876 the variable star name IL Aqr in {\it The 73rd Name List of
Variable Stars}. The Hipparcos satellite, however, made 67 brightness
measurements over the course of its three-year mission (Perryman et al. 1997)
and failed to detect photometric variability. These results are consistent with
the star's low level of chromospheric and coronal activity (Delfosse et al. 1998).

We have acquired high-precision photometric observations of GJ~876 with the T12
0.8~m automatic photometric telescope (APT) at Fairborn Observatory to measure
the level of photometric variability in the star, to derive its rotation
period, and to assess the possibility of observing planetary transits in the
GJ~876 system.  The T12 APT is equipped with a two-channel precision photometer
employing two EMI 9124QB bi-alkali photomultiplier tubes to make simultaneous
measurements in the Str\"omgren $b$ and $y$ passbands.  This telescope and its
photometer are essentially identical to the T8 0.8~m APT and photometer
described in Henry (1999).  The APT measures the difference in brightness
between a program star ($P$) and a nearby constant comparison star with a
typical precision of 0.0015 mag for bright stars ($V<8.0$). For GJ~876, we used
HD~216018 ($V=7.62$, $B-V=0.354$) as our primary comparison star ($C1$) and
HD~218639 ($V=6.43$, $B-V=0.010$) as a secondary comparison star ($C2$).  We
reduced our Str\"omgren $b$ and $y$ differential magnitudes with nightly
extinction coefficients and transformed them to the Str\"omgren system with
yearly mean transformation coefficients.  Further information on the telescope,
photometer, observing procedures, and data reduction techniques employed with
the T12 APT can be found in Henry (1999) and Eaton, Henry, \& Fekel (2003).

From June 2002 to June 2005, the T12 APT acquired 371 differential measurements
of GJ~876 with respect to the $C1$ and $C2$ comparison stars.  To increase the
precision of these observations, we averaged our Str\"omgren $b$ and $y$
magnitudes into a single $(b+y)/2$ passband.  The standard deviation of the
$C1-C2$ differential magnitudes from their mean is 0.0030 mag, slightly larger
than the typical 0.0015 mag precision of the APT observations.  However, since
the declination of GJ~876 is $-14\arcdeg$, the APT observations are taken at a
relatively high air mass, degrading the photometric precision somewhat.
Periodogram analysis of the $C1-C2$ differential magnitudes does not reveal any
significant periodicity, indicating that both comparison stars are
photometrically constant.  However, the standard deviations of the GJ~876
differential magnitudes with respect to the two comparison stars, $P-C1$ and
$P-C2$, are 0.0111 and 0.0110 mag, respectively, indicating definite
variability in GJ~876.  Our three sets of 371 $(b+y)/2$ differential magnitudes
are given in Table~5.

\begin{figure}
\includegraphics[scale=0.83]{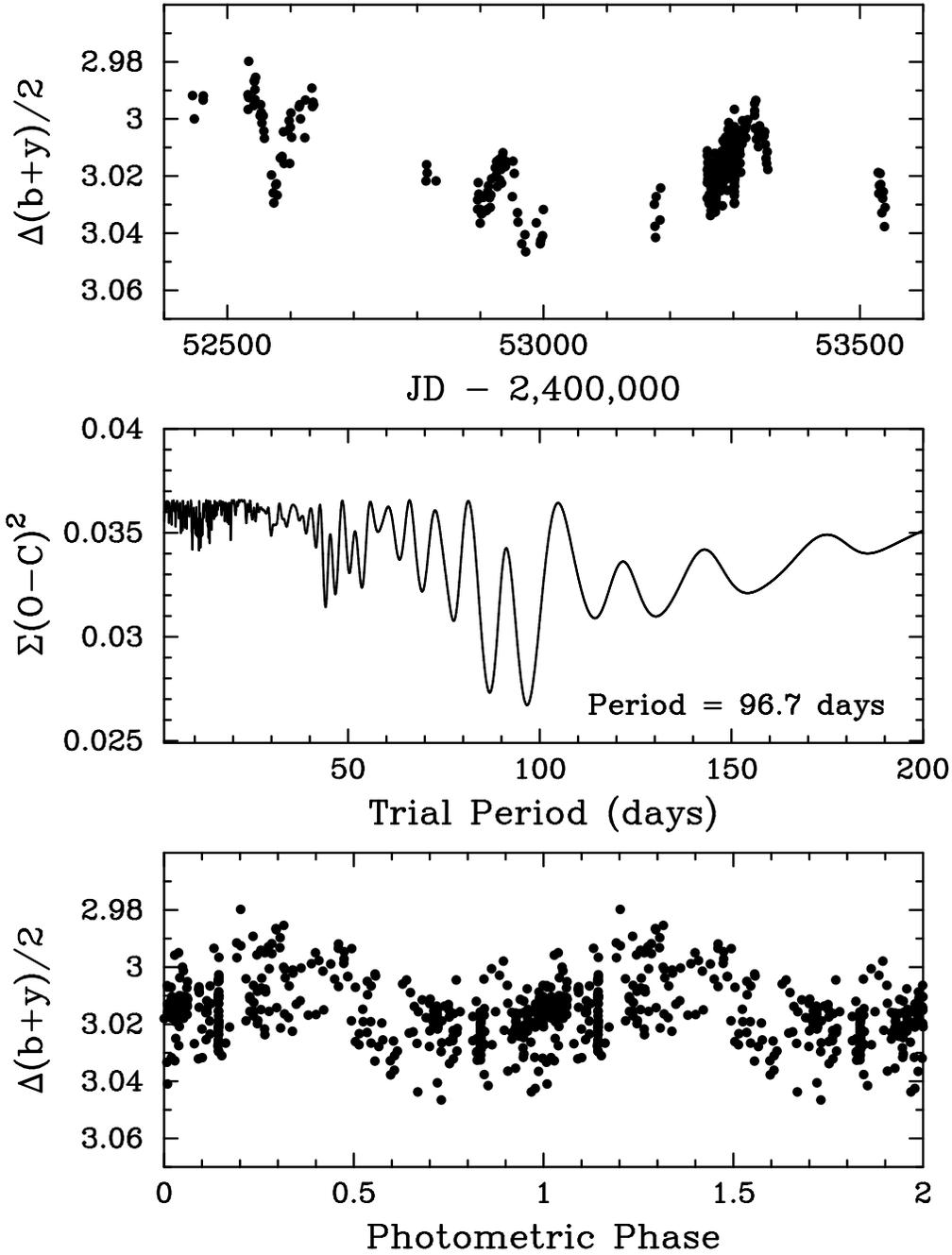}
\caption{
($Top$):  Photometric observations of GJ~876 with the T12 0.8~m
APT demonstrate variability of a few percent on a timescale of approximately
100 days along with longer-term variability. ($Middle$):  Periodogram
analysis of the APT observations gives the star's rotation period of
96.7 days. ($Bottom$):  The photometric observations phased with
the 96.7-day period reveal the effect of rotational modulation in the 
visibility of photospheric starspots on GJ~876.
}
\label{photometry_fig}
\end{figure}

The $P-C1$ differential magnitudes of GJ~876 are plotted in the top panel of
Figure \ref{photometry_fig}.  Photometric variability of a few percent is
clearly seen on a timescale of approximately 100 days; additional long-term
variability is present as well.  The light curve closely resembles those of
typical late-type stars with low to intermediate levels of chromospheric
activity (Henry, Fekel, \& Hall 1995).  Results of periodogram analysis of the
data in the top panel are shown in the middle panel, revealing a best period of
96.7 days with an estimated uncertainty of approximately one day.  We interpret
this period to be the rotation period of the star, made evident by modulation
in the visibility of photospheric starspots, which must cover at least a few
percent of the stellar surface.  The observations are replotted in the bottom
panel, phased with the 96.7-day period and an arbitrary epoch, and clearly
reveal the stellar rotation period.

There is little evidence from the photometric data for variability much shorter
than the rotation period of the star.  In particular, no photometric flares are
apparent. On two nights when GJ~876 was monitored for several hours (JD 2453271
and JD 2453301), the star appears to be constant to better than 1\%, consitent
with our measurement precision over a large range of airmass.  We conclude that
photometric transits of the planetary companions could be observed in this
star, if they occur, in spite of its intrinsic photometric variability.

\section{Photometric Limits on Transits by GJ~876 ``d''}
\label{transits}

The {\it a priori}  probability that a planet on a circular orbit transits its
parent star as seen from the line of sight to Earth is given by,
\begin{equation}
{\cal P}_{\rm transit}=0.0045 \left(\frac{1 {\rm AU}}{a}\right)\left(\frac{R_{\star}+R_{\rm pl}}{R_{\odot}}\right)
\end{equation}
where $a$ is the semi-major axis of the orbit and $R_{\star}$ and $R_{\rm pl}$
are the radii of the star and planet, respectively (Laughlin et al. 2005).  We
take $R_{\star}=0.3\,R_{\odot}$.  For a given composition, planetary density
increases with mass as higher pressures in the interior lead to greater
self-compression.  L\'{e}ger et al. (2004) find that the mean density of a
$6\,M_{\oplus}$ planet with composition similar to that of Earth would be
$\sim$ 39\% greater than that of our planet.  A $5.9\,M_{\oplus}$ planet with
such a density would have a radius of $1.619\,R_{\oplus}$, or
$0.0147\,R_{\odot}$.  Planets of comparable mass but containing significant
quantities of astrophysical ices and/or light gases
would be larger.  The third companion's orbital radius is $\sim$ 0.021 AU.
Thus, the {\it a priori} probability that the third companion transits GJ~876
is only $\sim7\%$.  The inclination of the orbit to the plane of the sky would
have to be $\gtrsim 86^{\circ}\, (\sim\arccos{0.07})$ to guarantee periodic
transits.  Until recently, radial velocity measurements provided little
constraint on the orbital inclinations of GJ~876's planets (Laughlin et al.
2005), and they still are only able to exclude a trivial fraction of possible
orientation angles for planet d (Section \ref{3plcf}).  Benedict et al. (2002)
astrometrically estimated companion b's inclination to the plane of the sky to
be $84\,\pm\,6^{\circ}$.  If we assume this range of possible values for the
system and that all three planets are nearly coplanar, the probability of a
transit by companion d rises to $\sim25\%$.  With a radius of
$1.619\,R_{\oplus}$, the transit depth is expected to be of order 0.22\%, which
is photometrically detectable by medium and large aperture telescopes.
Additionally, the transit duration can be as long as $(\arcsin{0.07}/\pi)P_d$,
or slightly over an hour.

We used previous N-body fits to generate model radial velocities, which were
then used to predict transit epochs for October, 2003.  We examined the reflex
radial velocity of the star due to just planet d; this motion is almost
periodic, as perturbations of planet d by planets b and c are small.  For a
planet on a circular orbit, transit centers should coincide with times when the
portion of the reflex velocity due to just the inner companion goes from
positive (red shifted) to negative (blue shifted).

Since the probability that planet d transits the face of its star is relatively
large, we obtained CCD photometry with the SITe\#3 camera (2048x3150 15 $\mu$m
pixels) at the Henrietta Swope 1 m telescope at Las Campanas, in an attempt to
detect such transits.  We observed for 6 consecutive nights (UT dates 10 to 15
Oct 2003), with possible transits expected (based on the RV ephemeris and the
1.94-day orbital period) during the nights of 10, 12, and 14 Oct.  We took all
observations using a Harris V filter; integration times were typically 100 s to
120 s, depending upon seeing and sky transparency.  With overheads, the
observing cadence was about 245 s per image; on a typical night we obtained
about 60 images, with a total of 355 usable images for the 6-night run.  The
nights of 10, 11, and 15 Oct were of photometric quality or nearly so; on 12,
13, and 14 Oct, each night began with an interval (roughly an hour long) of
thin to moderate cirrus over the target field.  Integration times were
necessarily long in order to maintain a moderate duty cycle, and to accumulate
enough total exposure time to reduce noise from atmospheric scintillation to
acceptable levels.  To avoid detector saturation for these relatively bright
stars, we therefore defocused the images to a width of about 30 CCD pixels.

Each CCD integration contained the image of GJ~876, as well as those of 10
other stars that were bright enough to use as comparison objects.  We computed
the extinction-corrected relative flux (normalized to unity when averaged over
the night of 10 Oct) of GJ~876 from the CCD images using proven techniques for
bright-star CCD photometry, as described by, e.g., Gilliland \& Brown (1992).
We removed residual drifts with typical amplitudes of 0.002 (which we attribute
to time-varying color-dependent extinction, combined with the extremely red
color of GJ~876) from the time series of GJ~876 by subtracting a quadratic
function of time (defined separately for each night).  After this correction,
the RMS variation of the relative flux for GJ~876 was in the range 0.001
to 0.0015 for each of the 6 nights.

We next searched for evidence of periodic transits by a small planet.  We did
this by folding the time series at each of a set of assumed periods, and then
convolving the folded series with negative-going boxcar functions (rectangular
window) with unit
depth and with widths of 30, 45, and 60 minutes.  We evaluated the convolution
with the boxcar displaced from the start of the folded time series by lag times
ranging from zero up to the assumed period, in steps of 0.005 day, or about 7
minutes. The convolution was normalized so that its value can be interpreted as
the depth (in relative flux units) of a transit-like signal with the same width
as the convolution boxcar function.  Our transit detection statistic consisted
of the normalized convolution, multiplied by the square root of the number of
data points lying within the non-zero part of the boxcar at any given value of
the lag.  For most periods and phases, the number of included data points is
about 15, so the detection statistic exceeds the transit depth by a factor of
about 4.  Since the expected duration of a transit by a short-period planet
across an M4 dwarf is about 60 minutes, the range of boxcar widths covers both
central and off-center transits, down to a duration for which the noise in the
convolution becomes prohibitive. We tested periods between 1.8 and 2.0 days, on
a dense period grid.

\begin{figure}
\includegraphics[scale=0.61]{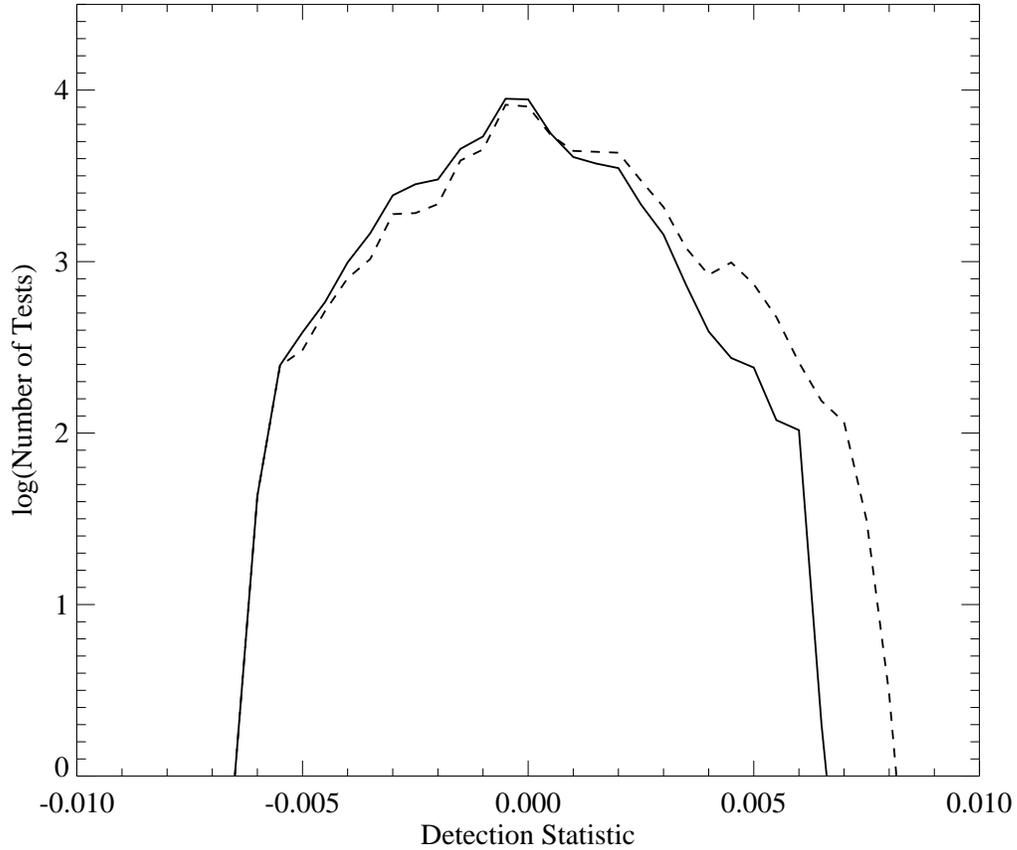}
\caption{
(Solid curve) Histogram of the logarithm, base 10, of the number of
period/phase combinations tested as a function of their resulting transit
detection statistics, for the full 6 nights of photometric data.  See text
for a definition of the transit detection statistic.  Note that this curve
is virtually symmetric about the value of 0, consistent with no transits
being observed.  (Dashed curve) Similar histogram for identical data to
which artificial transits with relative flux depth of 0.0015 have been
added, at a period of 1.900 days.
}
\label{Tim}
\end{figure}

The solid curve in Figure \ref{Tim} shows the logarithm of the histogram of the
detection statistic, computed using all of the data.  The largest transit-like
events that occur in the data set have detection statistics of 0.0068, but the
histogram is almost symmetrical about zero, so that there are very nearly as
many positive-going boxcar events as negative-going ones.  The value of the
transit amplitude for the planet's expected period and phase is 0.0005,
assuming a 60-minute transit; this is about 1.3 standard deviations larger than
zero.  From the distribution of transit amplitudes, we estimate the probability
that a true transit with amplitude 0.0015 is overlain by a negative-going noise
spike with amplitude $-0.001$ (yielding an observed signal of 0.0005) is only
about 2.4\%.  Thus, the observations contain no convincing evidence for
planetary transits within the period range searched, and within the range of
orbital phases probed by the data.

To refine our understanding of detectability, we added to the data synthetic
transits with various depths and 60-minute duration; the phases were chosen so
that transits occurred on each of the UT dates 10, 12, and 14 Oct.  The dashed
line in Figure \ref{Tim} shows the resulting histogram of the detection
statistic for synthetic transits with depth 0.0015.  This histogram is plainly
skewed toward positive values (negative-going transits), since real transits
produce not only a few very large values of the detection statistic, but also
many smaller ones (when the assumed period is not exactly correct, for
instance).  Examination of many realizations of synthetic transits suggests
that the skewness shown in Figure \ref{Tim} is near the limit of reliable
detection.  Adding synthetic transits with depths of 0.002, in contrast, always
produces a histogram that is unmistakably skewed. Accordingly, we conclude that
(within the period and phase limits already mentioned) there is no evidence for
a transiting planet that obscures more than 0.002 of the star's light, and it
is highly improbable that there are transits with depth as great as 0.0015.
Most likely, this is because the orbital inclination is such that transits do
not occur.  If transits are taking place, the maximum radius of the transiting
body is approximately $\sqrt{0.002}\,R_{\star} = 9.4 \times 10^3$ km, or
$1.47\,R_{\oplus}$, assuming the radius of GJ~876 to be $0.3\,R_{\odot}$.
Assuming a maximum transit depth of 0.0015, the corresponding planetary radius
is $1.28\,R_{\oplus}$. Note that a larger planet can have a non-central transit.

Even though companion d was not transiting in 2003, it may transit in the
future if the planets orbiting GJ~876 are on mutually inclined orbits.  An
analysis of the analogous case of possible transits by companion c is presented
by Laughlin et al. (2005).

\section{Discussion}
\label{dis}

The mass of GJ~876's third companion is $\sim7.5\,M_{\oplus}$. Assuming this
value of mass and a density of 8 g\,cm$^{-3}$ to account for a bit more
compression than that found for a $6\,M_{\oplus}$ rocky planet by L\'{e}ger et
al. (2004), the planet's radius is $1.73\,R_{\oplus}$. The escape velocity from
the surface would be slightly more than twice that of Earth, so that the planet
may well have retained a substantial atmosphere, and may thus have a larger
optical radius.

The proximity of GJ~876 d to its star implies that the timescale for tidal
synchronization of its rotation to its orbital period is short for any
reasonably assumed planetary properties.  However, it is possible that
atmospheric tides could maintain non-synchronous rotation, as they do on Venus
(Dobrovolskis 1980).  In analogy to models for Europa (Greenberg \&
Weidenschilling 1984), slightly non-synchronous rotation could result from an
eccentric orbit forced by perturbations from the outer planets, especially if
planet d lacks a substantial permanent asymmetry.  Note in this regard that the
topography of the planet's surface, if it has one, is likely to be muted as a
result of the high surface gravity ($\sim2.5$ times that of Earth) and the
expected malleability resulting from the planet's large potential for retaining
internal heat.  

The mean effective temperature of a planet orbiting at $a=0.021$ AU from a star
with $L=0.0124\,L_{\odot}$ is $T_{\rm effective} \sim 650 (1-A)^{1/4}$ K.
Assuming that heat is uniformly distributed around the planet, as it is on
Venus, and that the planet's albedo does not exceed 0.8, its effective
temperature should be in the range 430\,--\,650 K (157\,--\,377 C).
Simulations by Joshi et al. (1997), suggest that synchronously rotating
terrestrial planets with sufficiently massive atmospheres efficiently
redistribute heat to their unlit hemispheres.  For the opposite extreme of a
synchronously rotating planet with no redistribution of stellar heat, the
temperature at the subsolar point would be $\sqrt{2}$ higher at the substellar
point, and varies as the 1/4th power of the cosine of the angle between the
position of the star in the sky and the vertical (on the lit hemisphere),
implying very cold values near the terminator, and the unlit hemisphere would
be extremely cold. 

We can conceive of numerous possible scenarios to explain the
formation of GJ~876 d.  If the planet is rocky, it could have formed in
situ by accretion of small rocky planetesimals that spiraled inwards
as a result of angular momentum loss through interactions with gas
within the protoplanetary disk.  At the other extreme, GJ~876 d may be
the remnant core of a gas giant planet that migrated so close to its
star that it lost (most of) its gaseous envelope to Roche lobe
overflow (Trilling et al. 1998).  Neptune/Uranus-like formation
coupled with inwards migration models are also quite plausible, as
well as various other combinations of accretion/migration scenarios.
Additional observations of this and other small, close-in extrasolar
planets will be required to narrow down phase space enough for us to
have any confidence that any particular model is indeed the correct one.

The value of $\sqrt{\chi_{\nu}^2}$ = 1.065 for our effective 14+1 parameter fit
implies that the 3-planet coplanar inclined model provides an excellent fit to
the data. Nonetheless, the fit is not perfect, and additional variations may be
induced by stellar jitter and/or unmodelled small planets, as well as mutual
inclinations of the three known planets. We note that the residuals to both the
90$^{\circ}$ and the 50$^{\circ}$ 1.94-day 3-planet fits to the data have
modest power near 9 days (Figure \ref{3periodograms}); this period is also
present in many of the residuals to artificial data sets used to test various
aspects of the 3-planet fits.  Somewhat larger peaks near 13 and 120 days are
present in the residuals to the 50$^{\circ}$ fit.  A small planet with an
orbital period of around 13 days would be located so close to the massive and
eccentric planet c that it would not be stable for long unless it occupied a
protected dynamical niche.  Even around 9.4 days, long-term stability is
unlikely, especially if $i_c \lesssim 50^{\circ}$.  The peak at 120 days
probably represents an incomplete accounting of the radial velocity signature
of the two large planets, but it could also represent a small outer planet on a
resonant orbit.  We note that Ji, Li, \& Liu (2002, private communication) performed simulations of
the GJ~876 system with an additional planet with $a>0.5$ AU.  In
agreement with Rivera \& Lissauer (2001), all were found to be stable.  More
data are required to determine whether or not additional planets orbit GJ~876.

\section{Summary}
\label{sum}

We have shown that the GJ~876 system likely has a low mass planet on a close-in
orbit.  Fitting a model that includes the previously identified jovian-mass
companions to the radial velocity data obtained at the Keck telescope results
in residuals that contain significant power at a periodicity of 1.9379 days.
Including a third companion with this period in a self-consistent model results
in a significant improvement in the quality of the fit.  The third companion,
which we refer to as GJ~876 d, is found to have a minimum mass of
$5.89\,\pm\,0.54\,M_{\oplus}$ and an orbital period of 1.93776 $\pm$ 0.00007
days.  The corresponding semi-major axis is 0.021 AU, making it clearly the
smallest $a$ of any planet found in Doppler surveys.  Note that this is $\sim$
10 stellar radii, roughly coincident with the number of stellar radii
separating 51 Pegasi b from its host star.  Planet d is also probably closer to
its star in an absolute sense than are any of the planets found by transit.  A
significantly better fit to the data is obtained by assuming that the normal to
the three planets' orbits is inclined to the line of sight by $50^{\circ}$ than
by assuming this inclination to be $90^{\circ}$. For this $50^{\circ}$ fit, the
actual mass of the inner planet is $7.53\,\pm\,0.70\,M_{\oplus}$.

We have searched for transits, and find no evidence of them.  The lack of
observable transits means that we cannot place direct observational constraints
on planet d's radius and composition.  The requirement that the planet be
contained within its Roche lobe implies that its density is at least
0.075 g\,cm$^{-3}$, not a very meaningful bound.  Thus, while the newly
discovered companion may well be a giant rocky body, a large gaseous component
cannot be excluded.  Continued study of GJ~876 will provide us with additional
information on companion d, which may well be the most ``Earth-like'' planet
yet discovered.  See Table \ref{3plparami=50} for our best estimates of the
masses and orbital parameters of all three planets now known to orbit GJ~876.

\acknowledgments
We thank Drs. Ron Gilliland, Mark Phillips, and Miguel Roth for informative
advice and consultation during the observations performed at Las Campanas.  We
also are grateful for the contributed algorithms by Jason T. Wright, Chris
McCarthy, and John Johnson.  We thank Drs. Peter Bodenheimer, Eric Ford,
Man Hoi Lee, and Sara Seager for useful discussions.  We are also grateful to
Sandy Keiser for maintaining the computers used for this work at the Dept. of
Terrestrial Magnetism at the Carnegie Institution of Washington.  We thank an
anonymous referee for a thorough report which helped improve the paper.  The
work of EJR and JJL on this project was funded by NASA Solar Systems Origins
grant 188-07-21-03 (to JJL).  The work of GL was funded by NASA Grant NNG-04G191G
from the TPF Precursor Science Program.  We acknowledge support from the
Carnegie Institution of Washington and the NASA Astrobiology Institute through
Cooperative Agreement NNA04CC09A for EJR's work on photometric studies.  We
acknowledge support by NSF grant AST-0307493 and AST-9988087 (to SSV), travel
support from the Carnegie Institution of Washington (to RPB), NASA grant
NAG5-8299 and NSF grant AST95-20443 (to GWM), and by Sun Microsystems.  This
research has made use of the Simbad database, operated at CDS, Strasbourg,
France.  We thank the NASA and UC Telescope Time Assignment committees for
allocations of telescope time toward the planet search around M dwarfs.  GWH
acknowledges support from NASA grant NCC5-511 and NSF grant HRD-9706268.
Finally, the authors wish to extend a special thanks to those of Hawaiian
ancestry on whose sacred mountain of Mauna Kea we are very privileged to be
guests.  Without their generous hospitality, the Keck observations presented
herein would not have been possible.

\clearpage

\clearpage

\clearpage

\clearpage

\clearpage

\clearpage

\clearpage

\clearpage

\clearpage

\clearpage

\clearpage

\clearpage

\clearpage

\clearpage

\clearpage

\clearpage

\begin{deluxetable}{rrr}
\tablecaption{Measured Velocities for GJ~876 (Keck)}
\tablewidth{0pt}
\tablehead{
JD$~~$ & RV$~~$ & Unc.$~~$ \\
(-2450000)   &  (m\,s$^{-1}$) & (m\,s$^{-1}$)
}
\startdata
\tableline
 602.093 &  294.79 & 3.64 \\
 603.108 &  313.42 & 3.73 \\
 604.118 &  303.40 & 3.89 \\
 605.110 &  302.96 & 3.78 \\
 606.111 &  281.40 & 3.82 \\
 607.085 &  255.18 & 3.71 \\
 609.116 &  163.95 & 3.98 \\
 666.050 &  300.35 & 3.49 \\
 690.007 & -151.95 & 3.88 \\
 715.965 &  156.45 & 3.73 \\
 785.704 &  327.19 & 5.28 \\
 983.046 &  -96.43 & 3.61 \\
 984.094 & -117.02 & 4.00 \\
1010.045 &  -79.21 & 3.51 \\
1011.094 &  -62.54 & 3.70 \\
1011.108 &  -62.78 & 3.42 \\
1011.981 &  -35.71 & 3.12 \\
1011.989 &  -38.23 & 3.15 \\
1013.089 &   -7.85 & 3.49 \\
1013.963 &   14.53 & 3.73 \\
1013.968 &   16.75 & 3.80 \\
1043.020 &  -83.07 & 3.51 \\
1044.000 & -110.55 & 3.47 \\
1050.928 & -154.13 & 3.81 \\
1052.003 & -136.35 & 4.18 \\
1068.877 & -128.79 & 3.68 \\
1069.984 & -100.52 & 3.67 \\
1070.966 &  -94.38 & 3.60 \\
1071.878 &  -66.73 & 3.41 \\
1072.938 &  -55.33 & 3.69 \\
1170.704 & -123.56 & 4.41 \\
1171.692 & -137.08 & 4.02 \\
1172.703 & -119.17 & 4.04 \\
1173.701 & -115.95 & 4.35 \\
1312.127 & -134.51 & 3.37 \\
1313.117 & -133.52 & 3.79 \\
1343.041 &   35.17 & 3.80 \\
1368.001 & -182.55 & 4.00 \\
1369.002 & -191.05 & 4.14 \\
1370.060 & -174.57 & 3.51 \\
1372.059 & -157.56 & 5.86 \\
1409.987 &  -90.13 & 3.79 \\
1410.949 &  -85.59 & 3.86 \\
1411.922 &  -92.94 & 3.42 \\
1438.802 &  -63.41 & 3.90 \\
1543.702 & -142.50 & 4.02 \\
1550.702 & -197.70 & 3.84 \\
1704.103 &  122.99 & 3.76 \\
1706.108 &   73.75 & 3.91 \\
1755.980 &  272.62 & 5.08 \\
1757.038 &  245.87 & 4.24 \\
1792.822 & -215.71 & 3.74 \\
1883.725 &  187.77 & 3.96 \\
1897.682 &   50.01 & 4.27 \\
1898.706 &   42.34 & 4.14 \\
1899.724 &   32.04 & 3.53 \\
1900.704 &   13.98 & 3.47 \\
2063.099 &  212.65 & 4.01 \\
2095.024 & -228.99 & 4.35 \\
2098.051 & -266.92 & 4.71 \\
2099.095 & -257.23 & 4.51 \\
2100.066 & -270.35 & 3.92 \\
2101.991 & -248.41 & 3.84 \\
2128.915 &  130.58 & 5.15 \\
2133.018 &   55.95 & 4.35 \\
2133.882 &   68.55 & 4.86 \\
2160.896 & -269.15 & 3.84 \\
2161.862 & -270.68 & 4.28 \\
2162.880 & -235.10 & 4.49 \\
2188.909 &  116.79 & 4.39 \\
2189.808 &  113.52 & 5.00 \\
2236.694 &  187.17 & 4.10 \\
2238.696 &  208.12 & 4.08 \\
2242.713 &  225.32 & 4.95 \\
2446.071 &   84.52 & 4.99 \\
2486.913 &  194.04 & 4.46 \\
2486.920 &  195.16 & 4.46 \\
2487.120 &  182.36 & 4.16 \\
2487.127 &  181.67 & 4.06 \\
2487.914 &  179.58 & 4.53 \\
2487.923 &  180.93 & 4.48 \\
2488.124 &  188.92 & 3.98 \\
2488.131 &  181.99 & 4.15 \\
2488.934 &  162.66 & 3.75 \\
2488.940 &  162.29 & 3.69 \\
2488.948 &  161.35 & 3.50 \\
2488.955 &  163.13 & 3.52 \\
2514.867 & -121.06 & 4.78 \\
2515.873 & -143.91 & 4.27 \\
2535.774 &   45.38 & 4.43 \\
2536.024 &   49.11 & 3.87 \\
2536.804 &   77.18 & 5.01 \\
2537.013 &   75.82 & 4.12 \\
2537.812 &   87.35 & 4.09 \\
2538.014 &   91.81 & 4.48 \\
2538.801 &  121.30 & 4.42 \\
2539.921 &  137.37 & 3.92 \\
2572.709 &  -46.62 & 4.91 \\
2572.716 &  -44.75 & 5.20 \\
2572.916 &  -54.73 & 4.85 \\
2572.924 &  -50.10 & 5.60 \\
2573.740 &  -69.21 & 4.86 \\
2573.746 &  -66.75 & 4.80 \\
2573.875 &  -69.37 & 4.68 \\
2573.882 &  -66.53 & 4.70 \\
2574.760 & -104.45 & 4.41 \\
2574.768 & -102.01 & 4.53 \\
2574.936 & -103.60 & 4.85 \\
2574.944 & -102.76 & 4.91 \\
2575.716 & -124.28 & 4.80 \\
2575.722 & -123.40 & 4.16 \\
2600.748 &  134.05 & 3.95 \\
2600.755 &  134.67 & 3.92 \\
2601.747 &  138.56 & 4.05 \\
2601.754 &  141.02 & 4.35 \\
2602.717 &  160.52 & 4.50 \\
2602.724 &  164.31 & 4.77 \\
2651.718 & -116.96 & 6.23 \\
2807.028 &  168.61 & 4.21 \\
2829.008 & -240.74 & 4.06 \\
2832.080 & -170.81 & 4.35 \\
2833.963 & -121.76 & 4.07 \\
2835.085 &  -85.07 & 3.75 \\
2848.999 &  149.86 & 5.22 \\
2850.001 &  133.34 & 4.35 \\
2851.057 &  131.25 & 4.89 \\
2854.007 &   93.08 & 4.24 \\
2856.016 &  120.42 & 4.22 \\
2897.826 &  -42.50 & 4.05 \\
2898.815 &   -8.85 & 4.16 \\
2924.795 &  215.18 & 4.84 \\
2987.716 &  209.59 & 6.10 \\
2988.724 &  203.06 & 4.75 \\
3154.117 &   61.47 & 4.10 \\
3181.005 &  -51.18 & 4.51 \\
3181.116 &  -55.10 & 4.07 \\
3182.070 &  -93.33 & 4.00 \\
3191.037 & -257.70 & 4.19 \\
3195.970 & -173.67 & 4.25 \\
3196.997 & -156.17 & 4.64 \\
3301.808 &  -18.93 & 5.04 \\
3301.817 &  -25.17 & 5.55 \\
3301.823 &  -19.23 & 6.37 \\
3301.871 &  -25.66 & 4.18 \\
3302.723 &  -75.85 & 4.24 \\
3302.729 &  -72.85 & 4.43 \\
3302.736 &  -73.20 & 4.15 \\
3303.779 & -119.03 & 5.06 \\
3303.785 & -126.29 & 4.23 \\
3303.791 & -124.47 & 4.60 \\
3338.744 &   78.91 & 4.07 \\
3367.718 & -217.67 & 5.12 \\
3368.719 & -230.07 & 5.27 \\
3369.702 & -230.32 & 4.76 \\
3369.708 & -229.73 & 4.21 \\
\enddata
\label{velocities}
\end{deluxetable}

\newpage

\begin{deluxetable}{lll}
\tablecaption{Two-Planet $i=90^{\circ}$ Parameters}
\tablewidth{0pt}
\tablehead{
Parameter$~~$ & Planet c$~~$ & Planet b$~~$ \\
}
\startdata
\tableline
$m$\tablenotemark{a}      & 0.617 $\pm$ 0.007 $M_{\rm Jup}$ & 1.929 $\pm$ 0.009 $M_{\rm Jup}$\\
$P$ (days)                & 30.344 $\pm$ 0.018              & 60.935 $\pm$ 0.017\\
$K$ (m\,s$^{-1}$)         & 88.12 $\pm$ 0.94                & 212.04 $\pm$ 1.03\\
$a$\tablenotemark{a} (AU) & 0.13031 $\pm$ 0.00005           & 0.20781 $\pm$ 0.00004\\
$e$                       & 0.2232 $\pm$ 0.0018             & 0.0251 $\pm$ 0.0035\\
$\omega$ ($^{\circ}$)     & 198.3 $\pm$ 1.4                 & 176.8 $\pm$ 9.2\\
$M$ ($^{\circ}$)          & 308.8 $\pm$ 1.9                 & 174.3 $\pm$ 9.2\\
transit epoch             & JD 2452517.604 $\pm$ 0.067 & \\
\enddata
\tablenotetext{a}{Quoted uncertainties in planetary masses and semi-major axes {\it do not} incorporate the uncertainty in the mass of the star}
\label{2plparam}
\end{deluxetable}

\begin{deluxetable}{llll}
\tablecaption{Three-Planet $i=90^{\circ}$ Parameters}
\tablewidth{0pt}
\tablehead{
Parameter$~~$ & Planet d$~~$ & Planet c$~~$ & Planet b$~~$ \\
}
\startdata
\tableline
$m$\tablenotemark{a}      & 5.89 $\pm$ 0.54 $M_{\oplus}$ & 0.619 $\pm$ 0.005 $M_{\rm Jup}$ & 1.935 $\pm$ 0.007 $M_{\rm Jup}$\\
$P$ (d)                   & 1.93776 $\pm$ 0.00007        & 30.340 $\pm$ 0.013              & 60.940 $\pm$ 0.013\\
$K$ (m\,s$^{-1}$)         & 6.46 $\pm$ 0.59              & 88.36 $\pm$ 0.72                & 212.60 $\pm$ 0.76\\
$a$\tablenotemark{a} (AU) & 0.0208067 $\pm$ 0.0000005    & 0.13030 $\pm$ 0.00004           & 0.20783 $\pm$ 0.00003\\
$e$                       & 0 (fixed)                    & 0.2243 $\pm$ 0.0013             & 0.0249 $\pm$ 0.0026\\
$\omega$ ($^{\circ}$)     & 0 (fixed)                    & 198.3 $\pm$ 0.9                 & 175.7 $\pm$ 6.0\\
$M$ ($^{\circ}$)          & 309.5 $\pm$ 5.1              & 308.5 $\pm$ 1.4                 & 175.5 $\pm$ 6.0\\
transit epoch             & JD 2452490.756 $\pm$ 0.027   & JD 2452517.633 $\pm$ 0.051 & \\
\enddata
\tablenotetext{a}{Quoted uncertainties in planetary masses and semi-major axes {\it do not} incorporate the uncertainty in the mass of the star}
\label{3plparam}
\end{deluxetable}

\begin{deluxetable}{llll}
\tablecaption{Three-Planet $i=50^{\circ}$ Parameters}
\tablewidth{0pt}
\tablehead{
Parameter$~~$ & Planet d$~~$ & Planet c$~~$ & Planet b$~~$ \\
}
\startdata
\tableline
$m$\tablenotemark{a}      & 7.53 $\pm$ 0.70 $M_{\oplus}$ & 0.790 $\pm$ 0.006 $M_{\rm Jup}$ & 2.530 $\pm$ 0.008 $M_{\rm Jup}$\\
$P$ (d)                   & 1.93774 $\pm$ 0.00006        & 30.455 $\pm$ 0.019              & 60.830 $\pm$ 0.019\\
$K$ (m\,s$^{-1}$)         & 6.32 $\pm$ 0.59              & 87.14 $\pm$ 0.67                & 212.81 $\pm$ 0.66\\
$a$\tablenotemark{a} (AU) & 0.0208067 $\pm$ 0.0000004    & 0.13065 $\pm$ 0.00005           & 0.20774 $\pm$ 0.00004\\
$e$                       & 0 (fixed)                    & 0.2632 $\pm$ 0.0013             & 0.0338 $\pm$ 0.0025\\
$\omega$ ($^{\circ}$)     & 0 (fixed)                    & 197.4 $\pm$ 0.9                 & 185.5 $\pm$ 4.3\\
$M$ ($^{\circ}$)          & 311.8 $\pm$ 4.6              & 311.6 $\pm$ 1.3                 & 165.6 $\pm$ 4.2\\
\enddata
\tablenotetext{a}{Quoted uncertainties in planetary masses and semi-major axes {\it do not} incorporate the uncertainty in the mass of the star}
\label{3plparami=50}
\end{deluxetable}

\begin{deluxetable}{cccc}
\tablewidth{0pt}
\tablecaption{Photometric Observations of GJ~876}
\tablehead{
\colhead{Hel. Julian Date} & \colhead{$P-C1$} & \colhead{$P-C2$} & \colhead{$C1-
C2$} \\
\colhead{(HJD $-$ 2,400,000)} & \colhead{(mag)} & \colhead{(mag)} & \colhead{(mag)}
}
\startdata
52444.9553 & 2.9918 & 4.2587 & 1.2668 \\
52447.9492 & 3.0000 & 4.2730 & 1.2730 \\
52461.9151 & 2.9934 & 4.2622 & 1.2689 \\
52461.9541 & 2.9919 & 4.2609 & 1.2690 \\
52532.7211 & 2.9915 & 4.2610 & 1.2694 \\
\enddata
\tablecomments{Table 5 is presented in its entirety in the electronic edition
of the Astrophysical Journal.  A portion is shown here for guidance regarding
its form and content.}
\end{deluxetable}

\end{document}